\documentclass[nofootinbib,prd,showpacs]{revtex4-1}
\usepackage{amsmath}
\usepackage{amsfonts}
\usepackage{amssymb}
\usepackage{graphicx}
\graphicspath{ {figures/} }
\usepackage[usenames,dvipsnames]{xcolor}
\usepackage{hyperref}   
\hypersetup{colorlinks=true, 
citecolor=MidnightBlue, linkcolor=MidnightBlue, urlcolor=Cyan}
\usepackage{tikz}
\definecolor{purple}{rgb}{1,0,1}
\definecolor{lime}{HTML}{A6CE39} 

\newcommand{\orcidicon}{%
	\begin{tikzpicture}
	\draw[lime, fill=lime] (0,0) 
		circle [radius=0.16] 
		node[white] {{\fontfamily{qag}\selectfont \tiny ID}};
	\draw[white, fill=white] (-0.0625,0.095) 
		circle [radius=0.007];
	\end{tikzpicture}	\hspace{-2mm}
}
\newcommand\orcidFrancisco{{\href{https://orcid.org/0000-0002-9388-8373}{\orcidicon}}}
\newcommand\orcidMatt{{\href{https://orcid.org/0000-0003-1088-6485}{\orcidicon}}}
\newcommand\orcidAlex{{\href{https://orcid.org/0000-0002-1763-3563}{\orcidicon}}}
\newcommand\orcidManuel{{\href{https://orcid.org/0000-0001-8586-0285}{\orcidicon}}}
\newcommand\orcidMarcos{{\href{https://orcid.org/0000-0002-8080-9277}{\orcidicon}}}
\begin{document}
\title{
Novel black-bounce spacetimes: wormholes, regularity, energy conditions, and causal structure\\
}
\author{Francisco S. N. Lobo\orcidFrancisco\!\!}
\email{fslobo@fc.ul.pt}
\affiliation{Instituto de Astrof\'{i}sica e Ci\^{e}ncias do Espa\c{c}o, Faculdade de Ci\^encias da Universidade de Lisboa, Edif\'{i}cio C8, Campo Grande, P-1749-016, Lisbon, Portugal}
\author{Manuel E. Rodrigues\orcidManuel\!\!}
\email{esialg@gmail.com}
\affiliation{Faculdade de Ci\^{e}ncias Exatas e Tecnologia, 
Universidade Federal do Par\'{a}\\
Campus Universit\'{a}rio de Abaetetuba, 68440-000, Abaetetuba, Par\'{a}, 
Brazil}
\affiliation{Faculdade de F\'{\i}sica, Programa de P\'{o}s-Gradua\c{c}\~ao em 
F\'isica, Universidade Federal do 
 Par\'{a}, 66075-110, Bel\'{e}m, Par\'{a}, Brazil}
\author{Marcos V. de S. Silva\orcidMarcos\!\!}
\email{marco2s303@gmail.com}
\affiliation{Faculdade de F\'{\i}sica, Programa de P\'{o}s-Gradua\c{c}\~ao em 
F\'isica, Universidade Federal do 
 Par\'{a}, 66075-110, Bel\'{e}m, Par\'{a}, Brazil}
\author{Alex Simpson\orcidAlex\!\!}
\email{alex.simpson@sms.vuw.ac.nz}\affiliation{School of Mathematics and Statistics, Victoria University of Wellington,
PO Box 600, Wellington 6140, New Zealand}
\author{Matt Visser\orcidMatt\!\!}
\email{matt.visser@sms.vuw.ac.nz}\affiliation{School of Mathematics and Statistics, Victoria University of Wellington,
PO Box 600, Wellington 6140, New Zealand}
\date{Friday 25 September 2020; Monday 5 October 2020; \LaTeX-ed \today}
\begin{abstract}
We develop a number of novel ``black-bounce'' spacetimes. These are specific regular black holes where the ``area radius'' always remains non-zero, thereby leading to a ``throat'' that is either timelike (corresponding to a traversable wormhole), spacelike (corresponding to a ``bounce'' into a future universe), or null (corresponding to a ``one-way wormhole''). We shall first perform a general analysis of the regularity conditions for such a  spacetime, and then consider a number of specific examples. The examples are constructed using a mass function similar to that of Fan--Wang, and fall into several particular cases, such as the original Simpson--Visser model, a Bardeen-type model, and other generalizations thereof. We shall analyse the regularity, the energy conditions, and the causal structure of these models. The main results are several new geometries, more complex than before, with two or more horizons, with the possibility of an extremal case. We shall derive a general theorem regarding static space-time regularity, and another general theorem regarding (non)-satisfaction of the classical energy conditions. 

\bigskip

\end{abstract}
\pacs{04.50.Kd,04.70.Bw}
\maketitle
\def\HMS{{\scriptscriptstyle{\rm HMS}}}
\bigskip
\hrule
\tableofcontents
\bigskip
\hrule
\parindent0pt
\parskip7pt
\vspace{-10pt}
\section{Introduction}
\label{S:intro}
\vspace{-10pt}

Research in black hole physics has recently received a huge boost of interest, especially due to several breakthrough discoveries;  namely: (i) the reconstruction of the event-horizon-scale images of the supermassive black hole candidate in the centre of the giant elliptical galaxy M87 by the Event Horizon Telescope project~\cite{Akiyama:2019cqa, Akiyama:2019brx, Akiyama:2019sww,Akiyama:2019bqs,Akiyama:2019fyp,Akiyama:2019eap}; and (ii) the gravitational-wave searches by the LIGO Scientific and Virgo collaboration for coalescing compact binaries~\cite{Abbott:2016blz,Abbott:2017vtc,LIGOScientific:2018mvr,TheLIGOScientific:2016pea,TheLIGOScientific:2016wfe,TheLIGOScientific:2016htt,Abbott:2020khf} (and LISA in the future~\cite{LISA}). The detection of gravitational waves is not necessarily a completely definitive proof of the existence of black holes, since the ringdown signature in the time domain of extremely compact objects and black holes are very similar~\cite{Cardoso:2016rao}. Despite the fact that the exterior of a black hole is pathology-free, the interior seems to be riddled with problems~\cite{Cardoso:2019rvt}, such as the presence of spacetime singularities. More generically, the (maximally extended) Kerr family of solutions harbours closed timelike curves, and features Cauchy horizons signalling the breakdown of predictability of the theory~\cite{Penrosecosmic,Reallcosmic}. Nevertheless, as dictated theoretically by the weak cosmic censorship conjecture~\cite{Penrose:1969pc,Wald:1997wa}, spacetime singularities are cloaked by event horizons and therefore are inaccessible to distant observers.
In fact, there are still many subtle and interesting issues going on in black hole physics.  Deep issues of principle still remain, despite many decades of work on the subject, and in many cases it is worthwhile to carefully re-analyse and re-assess work from several decades ago~\cite{BH-in-GR,small-dark-heavy}. See also recent phenomenological discussions such as~\cite{observability,phenomenology,viability,complete,pandora,Doran:2006dq}. 

In particular, due to the problematic nature of the spacetime singularities, it is useful to consider the existence of regular black holes. It was recently shown that the spacetime structure of regular spherically symmetrical black holes generically entails the violation of the strong energy condition (SEC)~\cite{Zaslavskii}. 
In order to prove this, consider that, in general, there are $N$ zeros of the metric function $g_{tt}(r)$, located at the positions $r=u_i$ of the coordinate,  (where $i=1...N$). The outermost root viewed from the outside corresponds to the black hole event horizon.
In fact, it was shown that the SEC is violated in any static region within the event horizon in such a way that the Tolman mass becomes negative.  
In the non-static case, there is a constraint of another kind which, for a perfect fluid, entails the violation of the dominant energy condition (DEC).  

\clearpage
Furthermore, a general procedure for constructing exact regular black hole solutions was presented, in the presence of electric or magnetic charges in General Relativity (GR) coupled to a nonlinear electrodynamics (NLED)~\cite{NED4,Fan-Wang,Bronnikov-comment}. A two-parameter family of spherically symmetric black hole solutions were obtained, where the singularity at the spacetime centre was eliminated by moving to a certain region  in the parameter space; consequently the black hole solutions become regular everywhere. The global properties of the solutions were studied and the first law of thermodynamics was derived. The procedure was also generalized to include a cosmological constant, and regular black hole solutions that are asymptotic to an anti-de Sitter spacetime were constructed.

The study of regular black holes was generalized to modified theories of gravity and their relation with the energy conditions~\cite{Capozziello:2013vna,Capozziello:2014bqa}. For instance, a class of regular black hole solutions was obtained in four-dimensional $f(R)$ gravity, where $R$ is the curvature scalar, coupled to a nonlinear electromagnetic source~\cite{Rodrigues:2015}. Using the metric formalism and assuming static and spherically symmetric spacetimes, the resulting $f(R)$ and NLED functions were characterized by a one-parameter family of solutions which are generalizations of known regular black holes in GR coupled to NLED~\cite{NED1,NED2,NED3, NED4,NED5,NED6, Hollenstein:2008hp, NED7,NED8,NED9,NED10,Guerrero:2020uhn}. The related regular black holes of GR were recovered when the free parameter vanished, and where consequently the Einstein--Hilbert action was recovered, i.e., $f(R) \propto R$. The regularity of the solutions was further analysed and it was shown that there are particular solutions that violate only the SEC, which is consistent with the results attained in~\cite{Zaslavskii}.

This analysis was then generalized by leaving unspecified the function $f(R)$ and the NLED Lagrangian in the model, and regular solutions were then constructed through an appropriate  choice of the mass function~\cite{Rodrigues:2016}. It was shown that these solutions have two horizons, namely, an event horizon and a Cauchy horizon. All energy conditions are satisfied throughout the spacetime, except the SEC, which is violated near the Cauchy horizon. 
Regular solutions of GR coupled with NLED were also found by considering general mass functions and then  imposing the constraint  that the weak energy condition (WEC) and the DEC are simultaneously satisfied~\cite{Rodrigues:2017}. Further solutions of regular black holes were found by considering both magnetic and electric sources~\cite{Rodrigues:2018}, or adding rotation~\cite{bambi,neves,toshmatov,azreg,DYM,ramon}, or by considering  modified gravity~\cite{berej,rodrigues1,Silva:2018,Rodrigues:2019,Junior:2020,Cano:2020ezi}.

Herein, we are essentially interested in constructing regular black hole models, inspired by the recently developed ``black-bounce'' spacetimes~\cite{Simpson:2018tsi}. The constructed spacetime neatly interpolates between the standard Schwarzschild black hole and the Morris--Thorne traversable  wormhole~\cite{Morris:1988cz,Morris:1988tu,Visser:1995cc,Lobo:2017oab}, and at intermediate stages passes through a black-bounce, an extremal null-bounce, and a traversable wormhole. It is interesting to note that as long as the ``bounce'' parameter $a$ is non-zero the geometry is regular everywhere, so one has a somewhat unusual form of a ``regular black hole'', where $r = 0$ can be either spacelike, null, or timelike. Thus this spacetime generalizes and broadens the class of ``regular black holes'' beyond those usually considered. 

The non-static evolving version of this regular ``black-bounce'' geometry was also recently analysed, where the static metric was rewritten using Eddington-Finkelstein coordinates~\cite{Simpson:2019cer}.
In fact, the spacetime interpolates between the Vaidya spacetime, a black-bounce, and a traversable wormhole. It was also shown that the spacetime metric can be used to describe several physical situations of particular interest, including a growing black-bounce, a wormhole to black-bounce transition, and the opposite black-bounce to wormhole transition.
Furthermore, the black-bounce spacetimes were also used to construct closely related spherically symmetric thin-shell traversable wormholes, where each bulk region is now a segment of the black-bounce spacetime, and the exotic matter is concentrated on the thin shell~\cite{Lobo:2020kxn}. The construction permitted a dynamical analysis of the throat by considering linearized radial perturbations around static solutions, and it was shown that the stability of the wormhole is equivalent to choosing suitable properties for the exotic material residing on the wormhole throat.

It is interesting to note that different kinds of regular black holes to the ``black bounce'' solutions discussed above, with a minimum of the areal radius in the T-region, where the radial coordinate is timelike, or on a horizon, were discussed in \cite{K1,K2,K3,K4}.
More specifically, the spacetimes described in \cite{K2} have a de Sitter late-time asymptotic, making them in principle viable candidate cosmologies. In \cite{K4}, regular solutions with a phantom scalar and an electromagnetic field were obtained, leading to a diversity of global structures, including those with up to 4 horizons. In addition to this, the stability of the solutions obtained in \cite{K2} was analyzed in \cite{K5}, where it was shown that all the configurations under study were unstable under spherically symmetric perturbations, except for a special class of black universes where the event horizon coincides with the minimum of the area function.

Thus, in this work, we shall develop a number of additional novel ``black-bounce'' spacetimes. These are specific regular black holes where the ``area radius'' always remains non-zero, thereby leading to a ``throat'' that is either timelike (corresponding to a traversable wormhole), spacelike (corresponding to a ``bounce'' into a future universe), or null (corresponding to a ``one-way wormhole''). We shall first perform a general analysis of the regularity conditions for such a  spacetime, and then consider a number of specific examples.

The structure of this article is organized as follows. In section~\ref{S:general} we deal with general properties, like the spacetime symmetry and the curvature singularities in the subsection~\ref{SS:metric}; the stress-energy tensor~\ref{SS:stress-energy}; the Hernandez--Misner--Sharp mass in the subsection~\ref{SS:quasilocal-mass}; and the energy conditions in~\ref{SS:ECs}. In section~\ref{S:SV-bb} we describe some of the main features of the Simpson-Visser model. In section~\ref{S:new} we present several new black-bounce models, where we analyze the main characteristics; such as regularity, quasi-local mass, energy conditions and causal structure. We make our conclusion and final remarks in section~\ref{S:conclusion}.  
\enlargethispage{40pt}

We adopt the metric signature $(+,-,-,-)$. Given the Levi-Civita connection, $\Gamma^{\alpha}{}_{\mu\nu}=\frac{1}{2} g^{\alpha\beta}\left(\partial_{\mu}g_{\nu\beta}+\partial_{\nu}g_{\mu\beta}-\partial_{\beta}g_{\mu\nu}\right)$, the Riemann tensor is defined as $R^{\alpha}{}_{\beta\mu\nu}=\partial_{\mu}\Gamma^{\alpha}{}_{\beta\nu}-\partial_{\nu}\Gamma^{\alpha}{}_{\beta\mu}+\Gamma^{\sigma}{}_{\beta\nu}\Gamma^{\alpha}{}_{\sigma\mu}-\Gamma^{\sigma}{}_{\beta\mu}\Gamma^{\alpha}{}_{\sigma\nu}$. We shall work in geometrodynamic units where $G=c=1$. 

\section{General black-bounce spacetimes}\label{S:general}

\subsection{Metric and curvature}\label{SS:metric}

The most general static spherically symmetric metric can always locally be cast into the form:
\begin{eqnarray}
ds^2=f(r)\,dt^2-{dr^2\over f(r)} -\Sigma^2(r)\left(d \theta^2+\sin^2\theta d\phi^2\right)\label{ele}\,.
\end{eqnarray}
Here $f(r)$ and $\Sigma(r)$ are at this stage two freely specifiable functions. 
Horizons (if present) are located at the roots of $f(r)=0$, and the metric determinant is $g=-\Sigma^4(r) \sin^2\theta$. The area of a sphere at radial coordinate $r$ is $A(r) = 4\pi \Sigma^2(r)$. 
The coordinate choices implicit in equation \eqref{ele} are often called ``Buchdahl coordinates''~\cite{finch-skea,petarpa1,petarpa2,Semiz:2020}. 

From this line element, we may easily calculate the nonzero components of the Riemann tensor
\begin{eqnarray}
R^{tr}{}_{tr}=\frac{1}{2}f'', \qquad
R^{t\theta}{}_{t\theta}=R^{t\phi}{}_{t\phi}=\frac{f'\Sigma'}{2\Sigma},\qquad
R^{r\theta}{}_{r\theta}=R^{r\phi}{}_{r\phi}=\frac{f'\Sigma'+2f\Sigma''}{2\Sigma},\qquad
R^{\theta\phi}{}_{\theta\phi}=\frac{f\Sigma'^2-1}{\Sigma^2}\,.
\label{Rort}
\end{eqnarray}
To guarantee that the spacetime is everywhere regular we demand that:
\begin{itemize}
\itemsep-3pt
\item $\Sigma(r)$ must be non-zero everywhere.
\item $\Sigma'(r)$ and $\Sigma''(r)$ must be finite everywhere.
\item $f(r)$, $f'(r)$, and $f''(r)$ must be finite everywhere.
\end{itemize}
We may also calculate the Kretschmann scalar, $K=R_{\alpha\beta\mu\nu}R^{\alpha\beta\mu\nu}$, in terms of the Riemann components~\eqref{Rort},  as a semi-positive sum of squares~\cite{Bronnikov:2012wsj}
\begin{eqnarray}
&&K=4\left(R^{tr}{}_{tr}\right)^2+4\left(R^{t\theta}{}_{t\theta}\right)^2+4\left(R^{t\phi}{}_{t\phi}\right)^2+4\left(R^{r\theta}{}_{r\theta}\right)^2+4\left(R^{r\phi}{}_{r\phi}\right)^2+4\left(R^{\theta\phi}{}_{\theta\phi}\right)^2.
\label{Kret1}
\end{eqnarray}
More explicitly, in view of the spherical symmetry, we have
\begin{eqnarray}
&&K=4\left(R^{tr}{}_{tr}\right)^2+8\left(R^{t\theta}{}_{t\theta}\right)^2
+8\left(R^{r\theta}{}_{r\theta}\right)^2+4\left(R^{\theta\phi}{}_{\theta\phi}\right)^2.
\label{Kret2}
\end{eqnarray}
See appendix~\ref{S:appendix} for a more comprehensive justification of the fact that the Kretschmann scalar is semi-positive for the strictly static region of any static spacetime.
Specifically, in the current situation we find the explicit sum of squares
\begin{eqnarray}
&&K=\frac{
(\Sigma^2 f'')^2+2(\Sigma f' \Sigma')^2 +2\Sigma^2(f' \Sigma'+2f \Sigma'')^2
+4(1-f\Sigma'^2)^2}
{\Sigma^4}\,.
\label{Kret3}
\end{eqnarray}
Verifying whether or not the Kretschmann scalar is finite for all values of the radial coordinate $r$ is a good check on the regularity of any static spacetime. 

Similarly one can consider the Weyl scalar $C_{\mu\nu\alpha\beta} C^{\mu\nu\alpha\beta}$, for which a minor variant of the argument in~\cite{Bronnikov:2012wsj} yields:
\begin{eqnarray}
&&C_{\mu\nu\alpha\beta} C^{\mu\nu\alpha\beta}
=4\left(C^{tr}{}_{tr}\right)^2+4\left(C^{t\theta}{}_{t\theta}\right)^2
+4\left(C^{t\phi}{}_{t\phi}\right)^2+4\left(C^{r\theta}{}_{r\theta}\right)^2
+4\left(C^{r\phi}{}_{r\phi}\right)^2+4\left(C^{\theta\phi}{}_{\theta\phi}\right)^2.
\label{Csq1}
\end{eqnarray}
In view of spherical symmetry this reduces to
\begin{eqnarray}
&&C_{\mu\nu\alpha\beta} C^{\mu\nu\alpha\beta}=4\left(C^{tr}{}_{tr}\right)^2+8\left(C^{t\theta}{}_{t\theta}\right)^2
+8\left(C^{r\theta}{}_{r\theta}\right)^2+4\left(C^{\theta\phi}{}_{\theta\phi}\right)^2.
\label{Csq2}
\end{eqnarray}
Indeed, explicit computation yields a perfect square
\begin{eqnarray}
&&C_{\mu\nu\alpha\beta} C^{\mu\nu\alpha\beta}= {1\over3} \left(f'' - {2 f'\Sigma'\over\Sigma} 
+{2f (\Sigma'^2-\Sigma\Sigma'')\over\Sigma^2} -{2\over\Sigma^2}   \right)^2.
\label{Csq3}
\end{eqnarray}
Verifying whether or not the Weyl scalar is finite for all values of the radial coordinate $r$ is a partial check on the regularity of any static spacetime. 

\subsection{Stress-energy tensor}\label{SS:stress-energy}

The Einstein field equations are given by
\begin{eqnarray}
R_{\mu\nu}-\frac{1}{2}g_{\mu\nu}{R}=\kappa^2 T_{\mu\nu}\,,\label{eqEinstein}
\end{eqnarray}
where $g_{\mu\nu}$ is the metric tensor, $R_{\mu\nu}=R^{\alpha}{}_{\mu\alpha\nu}$, ${R}=g^{\mu\nu}R_{\mu\nu}$, $T_{\mu\nu}$ the stress-energy tensor and $\kappa^2=8\pi G/c^4$. In this work we adopt geometrodynamic units $c=G=1$,  so $\kappa^2\to8\pi$, as mentioned in the Introduction. If we consider the matter sector as an anisotropic fluid, then in regions where the $t$ coordinate is timelike, ($f(r)>0$, for instance, in the domain of outer communication), the mixed components of the stress-energy tensor are given by
\begin{eqnarray}
T^{\mu}{}_{\nu}={\rm diag}\left[\rho,-p_1,-p_2,-p_2\right]\,,\label{EMT}
\end{eqnarray}
where $\rho$, $p_1$, and  $p_2$ are the energy density and the two principal pressures, respectively. Taking into account the line element~\eqref{ele}, the Einstein equations~\eqref{eqEinstein} provide the following stress-energy profile
\begin{eqnarray}
&&\rho=-\frac{\Sigma \left(f' \Sigma'+2 f \Sigma''\right)+f \Sigma'^2-1}{\kappa ^2 \Sigma^2}\label{density+}\,,\\
&&p_1=\frac{\Sigma f'\Sigma'+f \Sigma'^2-1}{\kappa ^2 \Sigma^2}\,,\label{p1+}\\
&&p_2=\frac{\Sigma f''+2 f' \Sigma'+2 f \Sigma''}{2 \kappa ^2 \Sigma}\label{p2+}\,.
\end{eqnarray}

However, in regions where the $t$ coordinate is spacelike, $f(r)<0$, we should set
\begin{eqnarray}
T^{\mu}{}_{\nu}={\rm diag}\left[-p_1,\rho,-p_2,-p_2\right]\,,\label{EMT2}
\end{eqnarray}
where $p_1$ is the principal pressure in the now spacelike $t$ direction.
Then in the sub-horizon regions where $t$ is spacelike
\begin{eqnarray}
&&\rho=-\frac{\Sigma f'\Sigma'+f \Sigma'^2-1}{\kappa ^2 \Sigma^2}\,,\label{density-}\\
&&p_1=\frac{\Sigma \left(f' \Sigma'+2 f \Sigma''\right)+f \Sigma'^2-1}{\kappa ^2 \Sigma^2}\label{p1-}\,,\\
&&p_2=\frac{\Sigma f''+2 f' \Sigma'+2 f \Sigma''}{2 \kappa ^2 \Sigma}\label{p2-}\,.
\end{eqnarray}
Furthermore, at any horizons that may be present, where $f(r)=0$, we have
\begin{eqnarray}
&&\rho = - p_1=
-\frac{\Sigma f'\Sigma'-1}{\kappa ^2 \Sigma^2}\,,\label{density0} 
\qquad\qquad
p_2=\frac{\Sigma f''+2 f' \Sigma'}{2 \kappa ^2 \Sigma}\label{p20}\,.
\end{eqnarray}
The on-horizon equality of $\rho = - p_1$ has been known for some time~\cite{dbh2,dbh1}, and physically is needed to ensure that $\rho$ is continuous as one crosses the horizon.

Finally, for the trace of the stress energy
\begin{equation}
T= T^\mu{}_\mu = \rho-p_1-2p_2 = 
-{\Sigma^2 f'' +4 \Sigma(\Sigma'' f + \Sigma' f')  + 2 (\Sigma')^2 f - 2
\over \kappa^2 \Sigma^2} 
\end{equation}
regardless of whether one is above or below any horizon that may be present.

To guarantee that the stress-energy is everywhere regular we demand that:
\begin{itemize}
\itemsep-3pt
\item $\Sigma(r)$ must be non-zero everywhere.
\item $\Sigma'(r)$ and $\Sigma''(r)$ must be finite everywhere.
\item $f(r)$, $f'(r)$, and $f''(r)$ must be finite everywhere.
\end{itemize}
(This is of course the same set of conditions as was required for the Riemann tensor to be everywhere regular.)

\subsection{Hernandez--Misner--Sharp quasi-local mass}\label{SS:quasilocal-mass}
The Hernandez--Misner--Sharp quasi-local mass~\cite{Hernandez:1966, Misner:1964, Maeda:2007, Nielsen:2008, Abreu:2010, Faraoni:2020} is most easily defined by inspecting the 
Riemann tensor component
\begin{eqnarray}
R^{\theta\phi}{}_{\theta\phi}=-\frac{2M_\HMS(r)}{\Sigma(r)^3}
= {f(r) \Sigma'(r)^2-1\over\Sigma(r)^2}
\,.\label{Rmass}
\end{eqnarray}
Then
\begin{eqnarray}
M_\HMS(r)=\frac{1}{2}\Sigma(r)\left\{1-f(r)\Sigma'(r)^2\right\}\label{mass}\,.
\end{eqnarray}
Using this, $f(r)$ can be written as
\begin{eqnarray}
f(r)=1-\frac{2M_\HMS(r)-\Sigma(r)\{1-\Sigma'(r)^2\}}{\Sigma(r)\Sigma'(r)^2}\label{f}.
\end{eqnarray}
It will be useful to redefine $f(r) $ as
\begin{eqnarray}
f(r)=1-\frac{2M(r)}{\Sigma(r)}\label{fm}.
\end{eqnarray}
But now $M(r)$ is simply a function appearing in the metric, it is no longer the quasi-local mass obtained by integrating the energy density over the volume contained by a surface of radius $r$. Explicitly
\begin{equation}
M(r) = \frac{M_\HMS(r)-{1\over2} \Sigma(r)\{1-\Sigma'(r)^2\}}{\Sigma'(r)^2};
\qquad\qquad
M_\HMS(r) =  M(r) \Sigma'(r)^2+ {1\over2} \Sigma(r)\{1-\Sigma'(r)^2\}.
\end{equation}
At any horizon that may be present, where $f(r_H)=0$, in view of \eqref{mass} and \eqref{fm} we have
\begin{equation}
M_\HMS(r_H) = M(r_H) = {\Sigma(r_H)\over2}.
\end{equation}
Finally, at any local extremum of $\Sigma(r)$ that may be present, (that is $\Sigma'(r_{ext})=0$, corresponding to a ``throat'', a ``bounce'', or an ``anti-throat''), in view of \eqref{mass} and \eqref{fm} we have
\begin{equation}
M_\HMS(r_{ext}) = {\Sigma(r_{ext})\over2}; \qquad\qquad   M(r_{ext}) = {\Sigma(r_{ext})\; \{1-f(r_{ext})\}\over2}.
\end{equation}

Either differentiating $M_\HMS(r)$, or substituting~\eqref{f} into the Einstein equations, we may obtain the Hernandez--Misner--Sharp quasi-local mass in terms of the stress-energy component $T^t{}_t(r)$:
\begin{eqnarray}
M_\HMS'(r) = 4\pi T^t{}_t(r) \;\Sigma(r)^2\, \Sigma'(r); 
\qquad\qquad
M_\HMS(r)=M_*+4\pi\int_{r_*}^r T^t{}_t(\bar r)\; \Sigma(\bar r)^2\Sigma'(\bar r) d\bar r\,. 
\label{HMSmass}
\end{eqnarray}
Note that while the Hernandez--Misner--Sharp quasi-local mass can be defined for arbitrary values of $r$, it really only has its normal physical interpretation in the region where the $t$ coordinate is timelike, where $T^t{}_t \to \rho$ and we have:
\begin{eqnarray}
M_\HMS'(r) = 4\pi \rho(r) \;\Sigma(r)^2\, \Sigma'(r); 
\qquad\qquad
M_\HMS(r)=M_\HMS(r_H) +4\pi\int_{r_H}^r\rho(\bar r)\Sigma(\bar r)^2\Sigma'(\bar r) d\bar r\,. 
\label{HMSmass2}
\end{eqnarray}
That is, the ``mass function'' $M(r)$ defined in~\eqref{fm} is not the energy contained within a surface of radius $r$;  we now see that it is the Hernandez--Misner--Sharp quasi-local mass $M_\HMS(r)$ that plays this role. 

Note that in the limit $\Sigma(r)\rightarrow r$ we recover the usual results
\begin{eqnarray}
M(r)= M_\HMS(r); \qquad 
f(r)=1-\frac{2M_\HMS(r)}{r};\qquad
R^{\theta\phi}{}_{\theta\phi}=-\frac{2M_\HMS(r)}{r^3}.
\end{eqnarray}

\subsection{Energy conditions}\label{SS:ECs}

The standard energy conditions of classical GR are (mostly) linear in the stress-energy tensor, and have clear physical interpretations in terms of geodesic focussing, but suffer the drawback that they are often violated by semi-classical quantum effects.
In contrast, it is possible to develop non-standard energy conditions that are intrinsically non-linear in the stress-energy tensor, and which exhibit much better well-controlled behaviour when semi-classical quantum effects are introduced, at the cost of a less direct applicability to geodesic focussing~\cite{Martin-Moruno:2013a, Martin-Moruno:2013b, EC-LNP, Martin-Moruno:2015}.
The energy conditions have also found significant usage in cosmological settings~\cite{Visser:1997-epoch1,Visser:1997-epoch2,EC-galaxy, cosmo-99, Barcelo:2002, Cattoen:2006, Cattoen:2007}, in ``gravastars''~\cite{Visser:2003ge, Cattoen:2005he, MartinMoruno:2011rm, Lobo:2015lbc, Lobo:2012dp, black-stars}, and in various wormhole-related constructions~\cite{Visser:1989kh, Visser:1989kg, Poisson:1995sv, Visser:2003yf,pramana,Dadhich:2001fu}.
The standard point-wise energy conditions~\cite{Visser:1995cc}  for the stress-energy tensor~\eqref{EMT} are given by the inequalities
\begin{eqnarray}
&&NEC_{1,2}=WEC_{1,2}=SEC_{1,2} 
\Longleftrightarrow \rho+p_{1,2}\geq 0,\label{Econd1} \\[1pt]
&&SEC_3 \Longleftrightarrow\rho+p_1+2p_2\geq 0,\label{Econd2}\\[1pt]
&&DEC_{1,2} \Longleftrightarrow \rho-|p_{1,2}|\geq 0 \Longleftrightarrow 
(\rho+p_{1,2}\geq 0) \hbox{ and } (\rho-p_{1,2}\geq 0),\label{Econd3}\\[1pt]
&&DEC_3=WEC_3 \Longleftrightarrow\rho\geq 0,\label{Econd4}
\end{eqnarray}
(This formulation has carefully been phrased to be true regardless of whether the $t$ coordinate is timelike or spacelike.)
We note that $DEC_{1,2} \Longleftrightarrow ( (NEC_{1,2}) \hbox{ and } (\rho-p_{1,2}\geq 0))$. 
Since we already want to enforce the $NEC$, for all practical purposes we might as well subsume part of the $DEC$ into the $NEC$ and simply replace $DEC_{1,2}\Longrightarrow \rho-p_{1,2}\geq 0$.

Inserting the results given in~\eqref{density+}-\eqref{p2+}, in regions where the $t$ coordinate is timelike we have
\begin{eqnarray}
&&NEC_{1}=WEC_1=SEC_1 \Longleftrightarrow
-\frac{2 f \Sigma''}{\kappa ^2 \Sigma}\geq 0,\label{cond1+}\\
&&NEC_2=WEC_2=SEC_2 \Longleftrightarrow
\frac{\Sigma^2 f''-2 f \left(\Sigma \Sigma''+(\Sigma')^2\right)+2}{2 \kappa ^2 \Sigma^2}\geq 0,\label{cond2+}\\
&&SEC_3 \Longleftrightarrow
\frac{\Sigma f''+2 f' \Sigma'}{\kappa ^2 \Sigma}\geq0,\label{cond3+}\\
&&DEC_{1} \Longrightarrow
{2\left(1 - f' \Sigma\Sigma' - f (\Sigma')^2 -f \Sigma\Sigma'' \right)
\over\kappa^2\Sigma^2}\geq 0,\label{cond4+}\\
&&DEC_2 \Longrightarrow
-\frac{\Sigma^2 f''+\Sigma \left(4 f' \Sigma'+6 f \Sigma''\right)+2 f \Sigma'^2-2}{2 \kappa ^2 \Sigma^2}\geq0, \label{cond5+}\\
&&DEC_3=WEC_3 \Longleftrightarrow
-\frac{\Sigma \left(f' \Sigma'+2 f \Sigma''\right)+f (\Sigma')^2-1}{\kappa ^2 \Sigma^2}\geq0. \label{cond6+}
\end{eqnarray}

Inserting the results given in~\eqref{density-}-\eqref{p2-}, in regions where the $t$ coordinate is spacelike we have
\begin{eqnarray}
&&NEC_{1}=WEC_1=SEC_1 \Longleftrightarrow
+\frac{2 f \Sigma''}{\kappa ^2 \Sigma}\geq 0,\label{cond1-}\\
&&NEC_2=WEC_2=SEC_2 \Longleftrightarrow
{ \Sigma^2 f'' - 2 (\Sigma')^2 f +2 \Sigma \Sigma'' f +2 \over2\kappa^2\Sigma^2}\geq 0,
\label{cond2-}\\
&&SEC_3 \Longleftrightarrow
{ \Sigma f''+ 2 \Sigma' f' + 4 \Sigma'' f \over\kappa^2\Sigma} \geq0,
\label{cond3-}\\
&&DEC_{1} \Longrightarrow
{2\left(1 - f' \Sigma\Sigma' - f (\Sigma')^2 -f \Sigma\Sigma'' \right)
\over\kappa^2\Sigma^2}\geq 0,
\label{cond4-}\\
&&DEC_2 \Longrightarrow
{-\Sigma^2 f'' - 2 \Sigma \Sigma'' f  -4 \Sigma\Sigma' f' -2 (\Sigma')^2 f +2
\over2\kappa^2\Sigma^2} \geq 0,
\label{cond5-}\\
&&DEC_3=WEC_3 \Longleftrightarrow
-{\Sigma \Sigma' f' + (\Sigma')^2 f - 1 
\over\kappa^2\Sigma^2} \geq 0.
\label{cond6-}
\end{eqnarray}

That is, \emph{independent of whether one is above or below the horizon}, we have
\begin{equation}
NEC_{1}=WEC_1=SEC_1 \Longleftrightarrow
-\frac{2 |f(r)| \;\Sigma''(r)}{\kappa ^2 \Sigma(r)}\geq 0,\label{cond1b}\\
\end{equation}
So as long as one is not exactly on any event horizon that might be present we must have $f(r)\neq0$. Also $\Sigma(r)>0$ everywhere.
So we easily verify that $NEC_1=WEC_1=SEC_1$ all exhibit negative values everywhere not exactly on  the event horizon whenever $\Sigma''(r)>0$.  Thus the NEC, and so \emph{all} of the  standard point-wise energy conditions, are violated for  black-bounce models whenever $\Sigma''(r)>0$.

\textbf{Theorem}: For any static anisotropic fluid sphere with line element as in~\eqref{ele}, all of the standard point-wise energy conditions are violated whenever $f(r)\neq 0$, $\Sigma(r)>0$, and $\Sigma''(r)>0$.

Unfortunately, apart from $NEC_1$ and $DEC_1$,  the other point-wise energy conditions do not transform nicely as one crosses any horizon that may be present.

We now intend to look for models 
with positive energy density $\rho$, at least (insofar as possible) satisfying $WEC_3$. In addition to this, we are also looking for models 
that have a richer causal structure than the original Simpson--Visser model~\cite{Simpson:2018tsi}.

To quantify the amount of exotic matter present in the regions where the $NEC$ is violated, we may apply a volume integral quantifier~\cite{Visser:2003yf, Dadhich:2004}. With respect to Buchdahl coordinates, $\Sigma(r)$ defines the appropriate formula for the surface area of the spherical hypersurfaces \emph{via} $A = 4\pi \Sigma(r)^2$. It follows that if the $NEC$ is violated when $r\in\left[r_{1},r_{2}\right]$, then the amount of exotic matter is quantified by the definite integral

\begin{eqnarray}\label{volumequantifier}
    \int_{r_{1}}^{r_{2}}\left(\rho+p_{r}\right)4\pi\Sigma^2 \, d\Sigma &=& \int_{r_{1}}^{r_{2}}-\frac{2\vert f\vert\Sigma^{''}}{\kappa^2\Sigma}4\pi\Sigma^2 \, d\Sigma = -\int_{r_{1}}^{r_{2}}\vert f\vert\Sigma^{''}\Sigma \, d\Sigma \nonumber \\
    && \nonumber \\
    &=& -\int_{r_{1}}^{r_{2}}\vert f\vert\Sigma^{''}\Sigma^{'}\Sigma \, dr = -\frac{1}{2}\int_{r_{1}}^{r_{2}} \vert f\vert\Sigma\left[\left(\Sigma^{'}\right)^{2}\right]^{'} \, dr \nonumber \\
    && \nonumber \\
    &=& -\frac{1}{2}\vert f\vert\Sigma\left(\Sigma^{'}\right)^{2}\Bigg\vert_{r_{1}}^{r_{2}}+\frac{1}{2}\int_{r_{1}}^{r_{2}}\left(\vert f\vert\Sigma\right)^{'}\left(\Sigma^{'}\right)^{2} \, dr \ .
\end{eqnarray}
Given a specific candidate spacetime, \emph{i.e.} explicit forms for $f(r)$ and $\Sigma(r)$, we may compute this integral and obtain the amount of required exotic matter.

\section{Simpson--Visser black-bounce spacetime}
\label{S:SV-bb}

The Simpson--Visser black-bounce model is a special case of~\eqref{ele}. 
Specifically, take
\begin{eqnarray}
\Sigma(r)=\sqrt{r^2+a^2},\qquad M(r)=m,\qquad f(r)=1-\frac{2m}{\sqrt{r^2+a^2}}\,.\label{Visser}
\end{eqnarray}
This model has several properties, which we now list: 
Since (for $a>0$) $\Sigma(r)$ is never zero, and is regular, and $f(r)$ is regular, then for $a>0$ 
this spacetime is everywhere regular.
If we take the limit $a\rightarrow 0$,
then the Schwarzschild solution is recovered. For different values of the constant $a$, we have the following situations: (a) for $0<a<2m$, there are two horizons, $r_{\pm}=\pm\sqrt{(2m)^2-a^2}$, where $r_{+}$ is positive and $r_{-}$ is negative. 
This is a regular black hole spacetime, with the core being a bounce located at $r=0$; 
(b) for $a=2m$, we have a wormhole with a throat located at $r=0$, being an extremal null throat. This throat can only be crossed from one region to another, i.e., so that the wormhole is only one-way traversable; (c) for $a>2m$, we have a wormhole with a two-way timelike throat at $r=0$.

We may also see that in the case with two horizons, $0<a<2m$, so $f(r=0)=[(a-2m)/a]<0$. That is, $f(r)$ is positive outside the horizons with metric signature $(+,-,-,-)$, zero on the horizons, and negative between the horizons with metric signature $(-,+,-,-)$. 

For the Kretschmann scalar we find the explicit sum of squares
\begin{eqnarray}
&&K=\frac{
4m^2(2r^2-a^2)^2 + 8m^2r^4 
+ 8  \left(mr^2-2ma^2+a^2\sqrt{r^2+a^2}\right)^2
+4\left(2mr^2+a^2\sqrt{r^2+a^2}\right)^2
}{\left(r^2+a^2\right)^5}\,.
\end{eqnarray}
Provided $a> 0$ this is manifestly finite for all values of $r$ and $m$.

The energy conditions for this spacetime, in the region where $t$ is timelike, are written as
\begin{eqnarray}
&&NEC_1\Longleftrightarrow
-\frac{2 a^2 \left(\sqrt{r^2+a^2}-2 m\right)}{\kappa ^2 \left(r^2+a^2\right)^{5/2}} 
\geq 0,\qquad 
NEC_2\Longleftrightarrow
\frac{3 a^2 m}{\kappa ^2 \left(r^2+a^2\right)^{5/2}} \geq 0,
\label{NEC1out}\\
&&
WEC_3\Longleftrightarrow
-\frac{a^2 \left(\sqrt{r^2+a^2}-4 m\right)}{\kappa ^2 \left(r^2+a^2\right)^{5/2}} \geq 0,
\qquad
SEC_3\Longleftrightarrow 
\frac{2 a^2 m}{\kappa ^2 \left(r^2+a^2\right)^{5/2}}\geq 0,  
\\
&&DEC_1\Longrightarrow\frac{4ma^2}{\kappa ^2 \left(r^2+a^2\right)^{5/2}}\geq0\,,\,
\qquad\qquad\quad
DEC_2\Longrightarrow-\frac{a^2 \left(2\sqrt{r^2+a^2}-5m\right)}{\kappa ^2    \left(r^2+a^2\right)^{5/2}}\geq0.
\end{eqnarray}
Starting with the regular black hole spacetime, wherein $a<2m$, we see that the $NEC_1=WEC_1=SEC_1$ is violated outside the horizons $r_{\pm}$; furthermore the $WEC_3$ and $DEC_2$ are violated when $|r|\gg r_{+}$. 
For a wormhole with a null throat, $a=2m$, the $NEC_1$ is violated for all values of $r$; the $WEC_3$ and $DEC_2$ are violated for $|r|\gg a$. 
Relative to the 2-way wormhole with $a>2m$, the $NEC_1$ and $WEC_3$ are violated for all values of the radial coordinate, while $DEC_2$ is violated for $|r|$ sufficiently large. 
The energy density is always negative for the last case. 
It is noteworthy that spherically symmetric regular black holes in GR coupled to NLED always violate $SEC_3$, however, this is not necessarily true for black-bounce spacetimes.
Furthermore, even if $SEC_3$ is satisfied, $SEC_1$ is certainly violated --- at best one has \emph{partial} satisfaction of \emph{some} of the energy conditions.

The energy conditions for this spacetime, in the region where $t$ is spacelike, (the existence of this region requires $a<2m$),  are written as
\begin{eqnarray}
&&NEC_1\Longleftrightarrow
\frac{2 a^2 \left(\sqrt{r^2+a^2}-2 m\right)}{\kappa ^2 \left(r^2+a^2\right)^{5/2}} 
\geq 0,\qquad 
NEC_2\Longleftrightarrow
\frac{(2\sqrt{r^2+a^2} -m) a^2}{\kappa ^2 \left(r^2+a^2\right)^{5/2}} \geq 0,
\\
&&
WEC_3\Longleftrightarrow
\frac{a^2 }{\kappa ^2 \left(r^2+a^2\right)^{2}} \geq 0,
\qquad\qquad\quad
SEC_3\Longleftrightarrow 
\frac{2 a^2(2\sqrt{r^2+a^2} -3m)}{\kappa ^2 \left(r^2+a^2\right)^{5/2}}\geq 0,  
\\
&&DEC_1\Longrightarrow\frac{4ma^2}{\kappa ^2 \left(r^2+a^2\right)^{5/2}}\geq0\,,\,
\qquad\qquad\quad
DEC_2\Longrightarrow\frac{a^2 m }{\kappa ^2    \left(r^2+a^2\right)^{5/2}}\geq0.
\end{eqnarray}
Below the horizon we have $\sqrt{r^2+a^2}<2m$, so that $NEC_1=WEC_1=SEC_1$ is certainly violated. 
This implies that below the horizon all of the usual point-wise energy conditions are violated.
Even though $WEC_3$ is satisfied below the horizon, $WEC_1$ is not --- at best one has \emph{partial} satisfaction of \emph{some} of the energy conditions.

We may apply the volume integral from Eq.~(\ref{volumequantifier}) to the Simpson--Visser spacetime to obtain the amount of exotic matter required. In the case where $a>2m$, we have no horizons, \emph{i.e.}, a traversable wormhole geometry. For this case we may simply integrate the expression for $NEC_{1}$ above horizons from Eq.~(\ref{NEC1out}), all the way from $0$ to $+\infty$,

\begin{equation}
    \int_{0}^{+\infty}\frac{2a^2\left(2m-\sqrt{r^2+a^2}\right)}{\kappa^2\left(r^2+a^2\right)^{\frac{5}{2}}}\, dV = \int_{0}^{+\infty}\frac{a^2r(2m-\sqrt{r^2+a^2})}{(r^2+a^2)^2}\, dr = m-a \ .
\end{equation}
Given $a>2m$, the amount of exotic matter present must therefore be strictly greater than $m$ in order to stabilize the wormhole throat.

In the case where horizons are present, \emph{i.e.}, when we have a regular black hole and $a\in(0,2m)$, we find the following for the amount of exotic matter \emph{inside} the horizon:

\begin{equation}
    \int_{0}^{r_{H}}\frac{2a^2(\sqrt{r^2+a^2}-2m)}{\kappa^2(r^2+a^2)^{\frac{5}{2}}}\, dV = \int_{0}^{\sqrt{\left(2m\right)^2-a^2}} \frac{a^2r(\sqrt{r^2+a^2}-2m)}{(r^2+a^2)^2}\, dr = -\frac{(a-2m)^2}{4m} \ ,
\end{equation}
and for the amount of exotic matter \emph{outside} the horizon

\begin{equation}
    \int_{r_{H}}^{+\infty} \frac{2a^2(2m-\sqrt{r^2+a^2})}{\kappa^2(r^2+a^2)^{\frac{5}{2}}}\, dV = \int_{\sqrt{\left(2m\right)^2-a^2}}^{+\infty}\frac{a^2r(2m-\sqrt{r^2+a^2})}{(r^2+a^2)^2}\, dr = -\frac{a^2}{4m} \ .
\end{equation}
In all cases the amount of exotic matter required is strictly finite.

We can easily calculate the Hernandez--Misner--Sharp mass~\eqref{mass} for this model:
\begin{eqnarray}
M_\HMS(r)=
\frac{a}{2}+\frac{\kappa^2}{2}\int_0^r T^t{}_t(r)\; r\sqrt{r^2+a^2}dr=\frac{mr^2}{r^2+a^2}+\frac{a^2}{2\sqrt{r^2+a^2}}
\label{massVisser}\,.
\end{eqnarray}

This mass is always positive. The first identity in this expression arises from equation~\eqref{HMSmass}, with $r_*=0$, $M_*=a/2$. We also have the limits $\lim_{r\rightarrow 0}M_\HMS(r)=a/2$ and $\lim_{r\rightarrow \infty}M_\HMS(r)=m$. 

\begin{figure}[!htb]
	\includegraphics[scale=0.40]{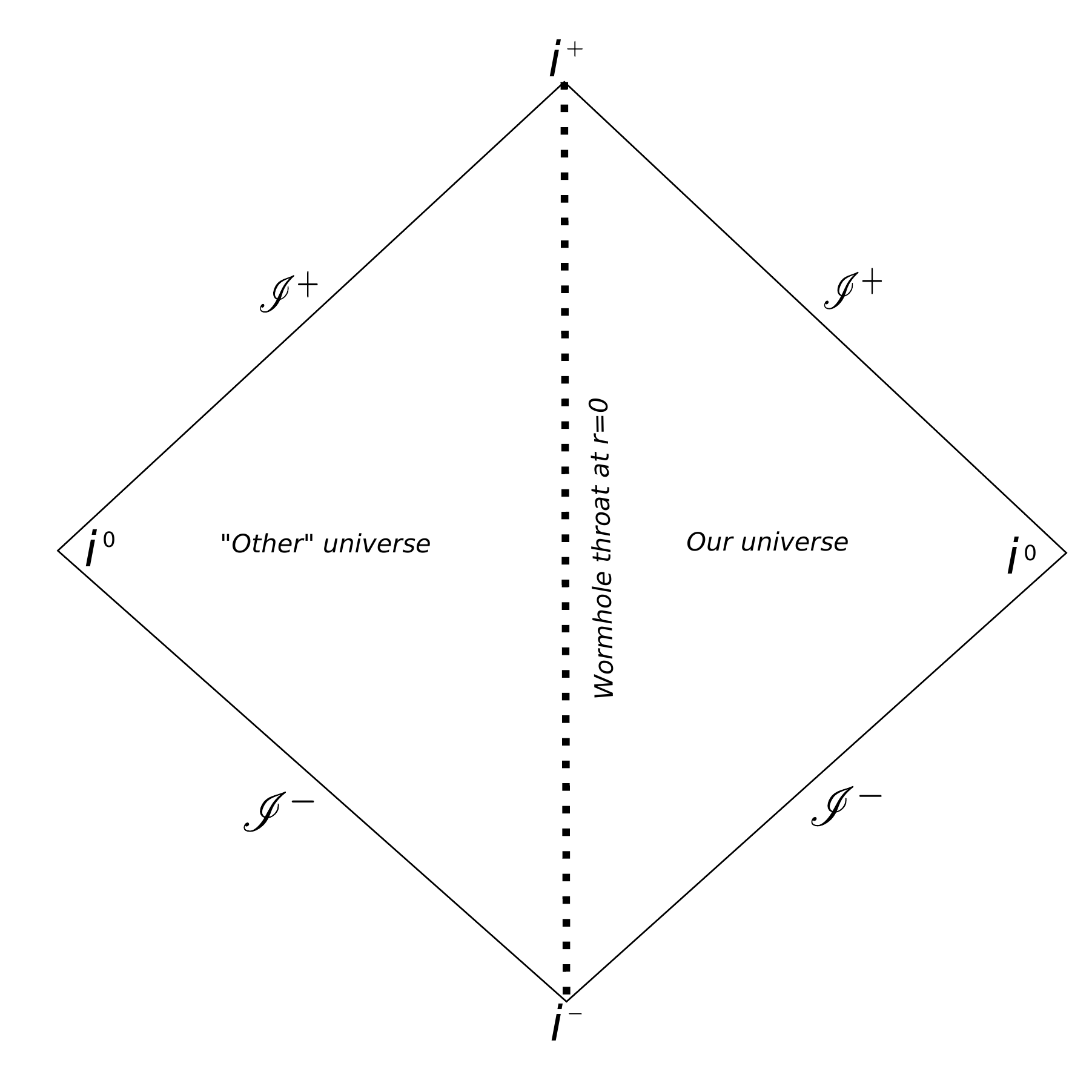}
	\caption{Carter--Penrose diagram for the case when we have a traditional two-way traversable wormhole in the sense of Morris and Thorne.}\label{figPenrose1}
\end{figure}

\begin{figure}[!htb]
	\includegraphics[scale=0.40]{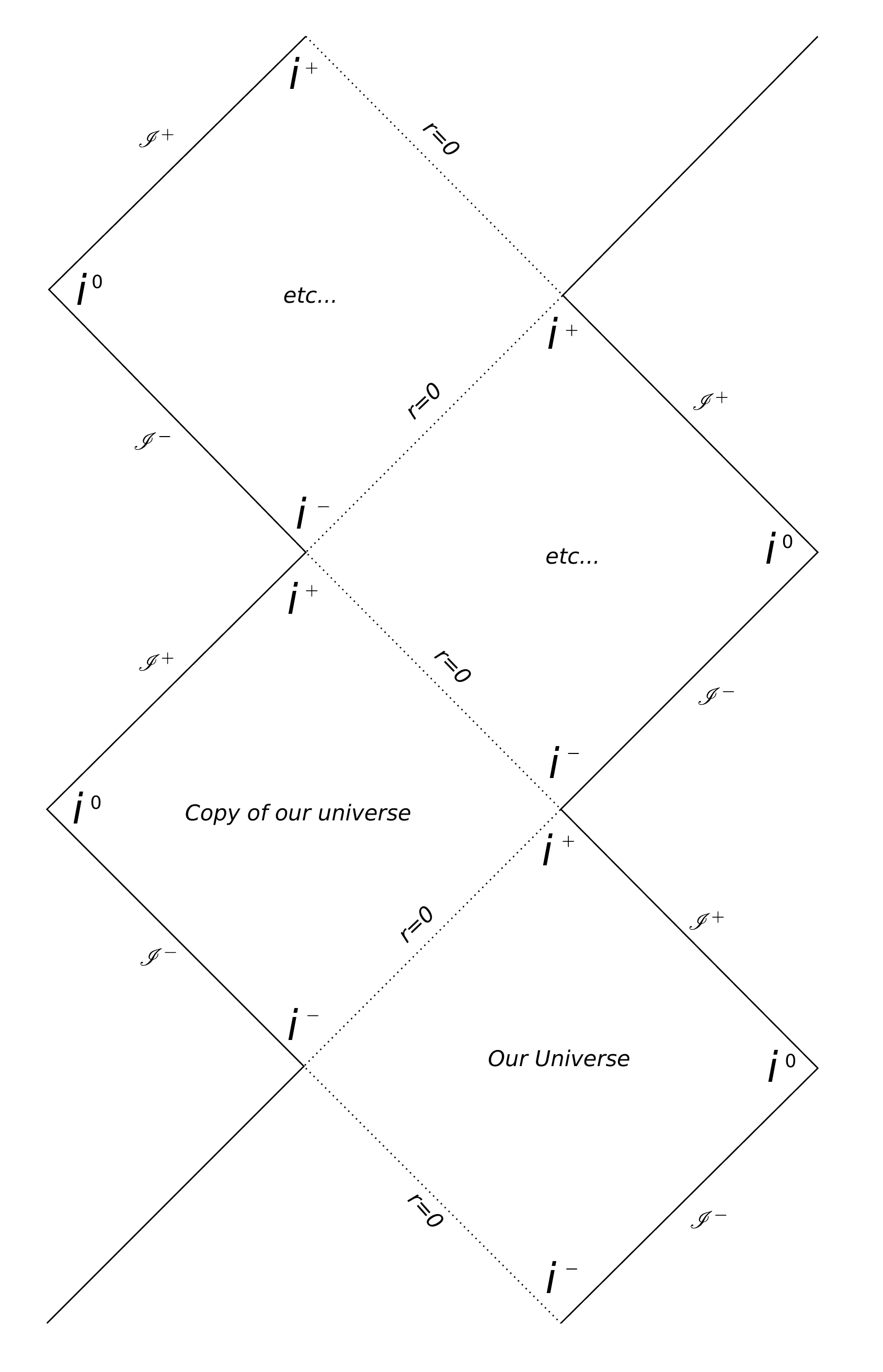}
	\caption{Carter--Penrose diagram for the case when we have a one–way wormhole geometry with an extremal null throat; for all relevant examples this corresponds to fixing $a = a_{ext}$.}\label{figPenrose2}
\end{figure}

\begin{figure}[!htb]
	\includegraphics[scale=0.45]{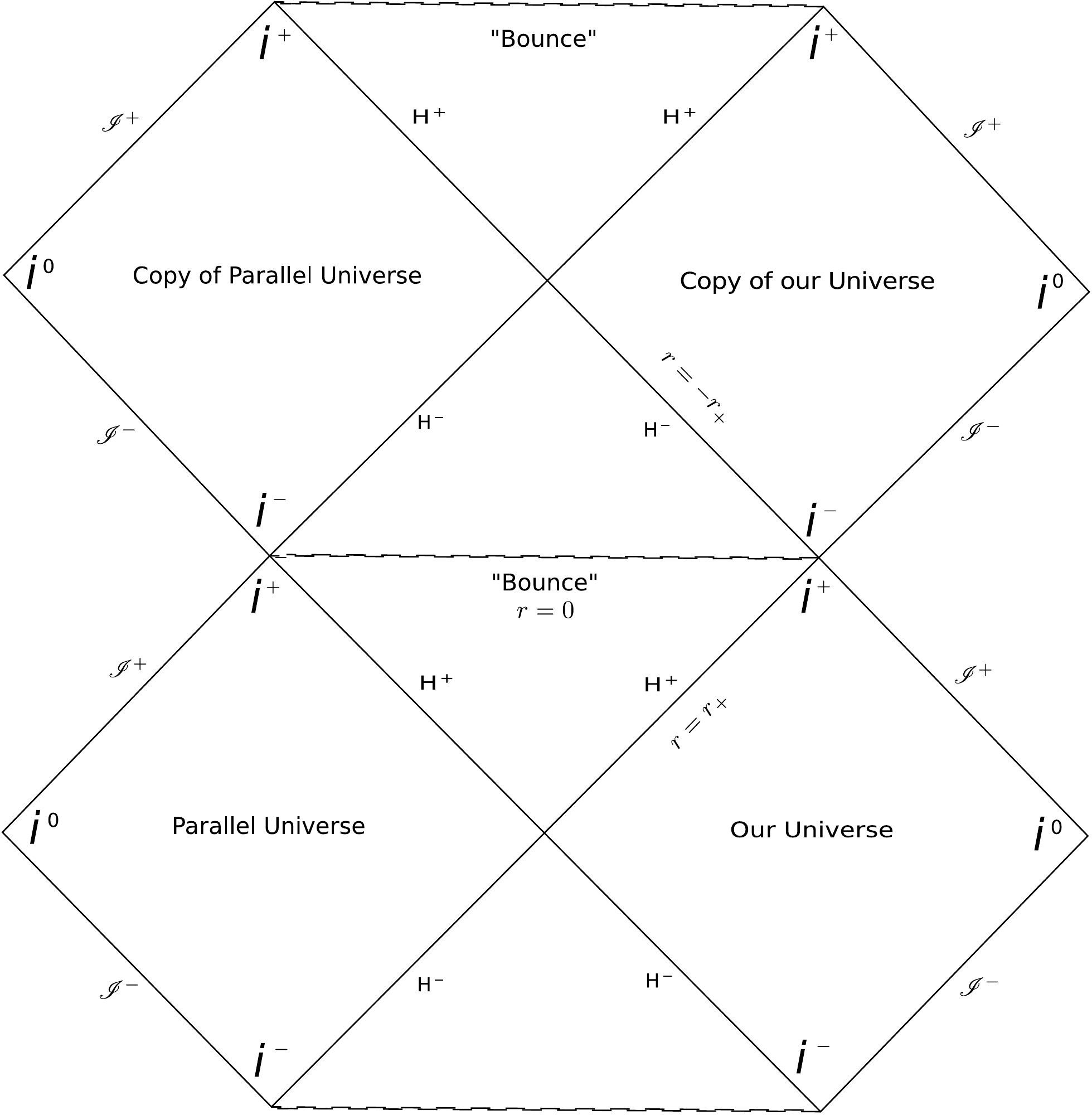}
	\caption{Carter--Penrose diagram for the maximally extended case where we have one horizon location in each universe. One may propagate through the event horizon, at $r = r_+$, to reach the spacelike ``bounce'' hypersurface at $r = 0$, before ``bouncing'' into a future reincarnation of our own universe. Infinitely many future copies of the universe exist if we extrapolate the time coordinate `up' the page; only two are displayed here for tractability.}\label{figPenrose3}
\end{figure}

The causal structure of the spacetime is given by the Carter--Penrose diagrams for the following cases:
(i) For $a > a_{ext} = 2m$, in  Fig.~\ref{figPenrose1}, which corresponds to a traditional two–way traversable wormhole in the sense of Morris and Thorne; (ii) for $a = a_{ext} = 2m$, in  Fig.~\ref{figPenrose2}, which corresponds to a one–way wormhole geometry with an extremal null throat; (iii) for $0 < a < 2m$, in  Fig.~\ref{figPenrose3}, where we have one horizon location in each universe, and  one may propagate through the event horizon, at $r = r_+$, to reach the spacelike ``bounce'' hypersurface at $r = 0$, before ``bouncing'' into a future reincarnation of our own universe.

\enlargethispage{50pt}

Our primary goal below is to explore new black-bounce models that generalize this Simpson--Visser model, or that might somewhat ameliorate the violation of the energy conditions. We shall explore these options in the next few sections. In the rest of the work we will consider $\Sigma(r)=\sqrt{r^2+a^2}$ as given in~\eqref{Visser}.

Though not central to this paper, a dynamical version of the Simpson--Visser spacetime has been explored in ~\cite{Simpson:2019cer}, where multiple phenomenological models describing various transitions are analysed.
 
\section{New black-bounce spacetimes}\label{S:new}

We shall first consider the following rather general class of black-bounce models that generalize the Simpson--Visser spacetime, in which the functions $\Sigma(r)$, $M(r)$ and $f(r)$ are given by
\begin{eqnarray}
\Sigma(r)=\sqrt{r^2+a^2}, \qquad
M(r)=\frac{m\Sigma(r)\, r^{k}}{\left(r^{2n}+a^{2n}\right)^{(k+1)/(2n)}},\qquad
 f(r)=1-\frac{2M(r)}{\Sigma(r)}\label{newBBS}\,.
\end{eqnarray}
Here $n$ and $k$ are positive integers. This new model is inspired by the Fan--Wang mass function~\cite{Fan-Wang} for regular black holes. The Simpson--Visser model~\eqref{Visser} is recovered for $n=1$ and $k=0$, and (for any $n$ and $k$) in the limit $a\rightarrow 0$ we obtain the Schwarzschild solution. We cannot recover the usual regular black hole solutions (Bardeen, Hayward, Frolov) due to the term $a^2$ present in $\Sigma(r)$. However, this model can generate several new classes of black-bounce, for which we shall examine several specific cases below.
\enlargethispage{60pt}

\subsection{Model $n=2$ and $k=0$}\label{SS:n=2+k=0}

For $n=2$ and $k=0$ in~\eqref{newBBS}, we have
\begin{eqnarray}
\Sigma(r)=\sqrt{r^2+a^2},\qquad f(r)=1-\frac{2m}{\sqrt[4]{r^4+a^4}}\label{mod1}\,.
\end{eqnarray}
In order to verify the regularity of the model, we analyse the Kretschmann scalar~\eqref{Kret1}, using~\eqref{mod1}, which takes the following form:
\begin{eqnarray}
K=\frac{8 m^2 r^8}{\left(r^2+a^2\right)^2 \left(r^4+a^4\right)^{5/2}}
+\frac{{4 m^2 r^4 \left(3 a^4 -2
   r^4\right)^2}}{ \left(r^4+a^4\right)^{9/2}}
   +\frac{4 \left(\frac{2 m
   r^2}{\sqrt[4]{a^4+r^4}}+a^2\right)^2}{\left(r^2+a^2\right)^4}
   		\nonumber\\
   +\frac{8 \left(a^2\left(r^4+a^4\right)^{5/4}+m\left(r^6-2a^6-a^2r^4\right)\right)^2}{\left(r^2+a^2\right)^{4} \left(r^4+a^4\right)^{5/2}}\,.
\end{eqnarray}

The Kretschmann scalar is manifestly finite for all real values of the radial coordinate, so the spacetime is regular for this model. From~\eqref{mod1} we see that $f(r)=0$ provides two symmetric real values $r_{\pm}=\pm\sqrt[4]{(2m)^4-a^4}$. When $0<a<2m$, we have a regular black hole with two event horizons $r_{\pm}$, one in the positive and another in the negative region of $r$, with signature $(+,-,-,-)$ outside the horizon. For $r=0$, the spacetime is regular and can be extended to $r<0$, then going through a bounce to the region where the radial coordinate is negative, i.e., this corresponds to a one-way spacelike throat, a ``black bounce''. We also see that the limit of $f(r)$ for $r\rightarrow 0$, results in $[(a-2m)/a]<0$, with a signature $(-,+,-,-)$ inside the horizon. If $a=2m$, the (maximally extended) spacetime has only extremal horizons, so we have a one-way wormhole geometry with an extremal null throat, and a signature $(+,-,-,-)$. For $a>2m$, we have no horizons and there is a wormhole with a two-way timelike throat, maintaining the signature $(+,-,-,-)$ throughout. Essentially, this spacetime possesses the same characteristics as for the Simpson--Visser geometry. Thus, this is the behaviour in general for any integer $n$  and  $k=0$.
Thus, the causal structures remain the same as in the Simpson--Visser case. More specifically, (i) the case $a>a_{ext} =2m$ is depicted in Fig.~\ref{figPenrose1}; (ii) $a=a_{ext} =2m$ in Fig.~\ref{figPenrose2}; (iii) and the region $0<a<a_{ex}=2m$ is depicted in Fig.~\ref{figPenrose3}.

In the region where the $t$ coordinate is timelike, the energy conditions are given by
\begin{eqnarray}
NEC_1\Longleftrightarrow -\frac{2 a^2 \left(1-\frac{2 m}{\sqrt[4]{a^4+r^4}}\right)}{\kappa ^2 \left(r^2+a^2\right)^2} \geq 0\,,\,
\qquad\qquad
NEC_2\Longleftrightarrow \frac{ma^2 \left(2 a^6+3 a^4 r^2+7 a^2 r^4-2  r^6\right)}{\kappa ^2 \left(r^2+a^2\right)
   \left(r^4+a^4\right)^{9/4}} \geq0  \,,
\end{eqnarray}   

\begin{equation}   
WEC_3\Longleftrightarrow -\frac{a^2 \left(\left(r^4+a^4\right)^{5/4}
-2m(2a^4+a^2r^2+ r^4)\right)}{\kappa ^2 \left(r^2+a^2\right)^2 \left(r^4+a^4\right)^{5/4}} \geq0 \,,
\qquad
SEC_3\Longleftrightarrow  \frac{2 m a^2 r^2\left(3 a^4 +5 a^2 r^2-2 r^4\right)}{\kappa ^2 \left(r^2+a^2\right)
   \left(r^4+a^4\right)^{9/4}}\geq 0\,,
 \end{equation}
 \begin{equation}  
DEC_1\Longrightarrow  \frac{4 a^4 m}{\kappa ^2 \left(r^2+a^2\right) \left(r^4+a^4\right)^{5/4}} \geq 0 \,,
\qquad\qquad  
   DEC_2\Longrightarrow \frac{a^2 \left(\frac{m \left(6 a^8-a^6 r^2+2 a^4 r^4-a^2 r^6+6 r^8\right)}{\left(r^4+a^4\right)^{9/4}}-2\right)}{\kappa
   	^2 \left(r^2+a^2\right)^2}\geq 0\,.
\end{equation}

We see that outside the horizons the $NEC_1=WEC_1=SEC_1$ condition is again violated for $|r|>r_{+}$ and the $NEC_2$, $SEC_3$, $DEC_{2}$ and $WEC_3$ are violated for $r\gg r_{+}$. Again, we have a negative energy density beyond some in-principle calculable but messy absolute value of $r$, maintaining the essential characteristics of the original Simpson--Visser model.

In the sub-horizon region where the $t$ coordinate is spacelike (the existence of this region requires $a<2m$) we have:
\begin{eqnarray}
NEC_1\Longleftrightarrow \frac{2 a^2 \left(1-\frac{2 m}{\sqrt[4]{a^4+r^4}}\right)}{\kappa ^2 \left(r^2+a^2\right)^2} \geq 0\,,\,
\end{eqnarray}   
\begin{eqnarray}   
NEC_2\Longleftrightarrow 
 \frac{{2} a^2 \left(r^4+a^4\right)^{9/4} - ma^2 (3r^2-a^2)\left(2r^6-r^4a^2-r^2a^4-{2}a^6\right)}{\kappa ^2 \left(r^2+a^2\right)^2
   \left(r^4+a^4\right)^{9/4}} \geq0  \,,
\end{eqnarray}   
\begin{equation}   
WEC_3\Longleftrightarrow \frac{a^2 \left(\left(r^4+a^4\right)^{5/4}
-2mr^2(r^2-a^2)\right)}{\kappa ^2 \left(r^2+a^2\right)^2 \left(r^4+a^4\right)^{5/4}} 
\geq0 \,,
\end{equation}
\begin{equation}  
SEC_3\Longleftrightarrow  \frac{4a^2\left(r^4+a^4\right)^{9/4} -2 m a^2\left(4 a^8 -3a^6r^2-3a^2r^6+6r^8\right)}{\kappa ^2 \left(r^2+a^2\right)^2
   \left(r^4+a^4\right)^{9/4}}\geq 0\,,
\end{equation}
\begin{equation}  
DEC_1\Longrightarrow  \frac{4 a^4 m}{\kappa ^2 \left(r^2+a^2\right) \left(r^4+a^4\right)^{5/4}} \geq 0 \,,
\qquad\qquad  
   DEC_2\Longrightarrow 
   {ma^2(2a^4-5r^2a^2+2r^4)\over \kappa^2 (r^4+a^4)^{9/4}}\geq 0\,.
\end{equation}

We see that between the horizons the $NEC_1=WEC_1=SEC_1$ condition is again violated, now  for all $|r|<r_{+}$.
This implies violation of all the standard point-wise energy conditions, maintaining the essential characteristics of the original Simpson--Visser model.

The Hernandez--Misner--Sharp mass~\eqref{mass} for the case of~\eqref{mod1} is given by
\begin{eqnarray}
M_\HMS(r)=\frac{a^2}{2\sqrt{r^2+a^2}}+\frac{mr^2}{\sqrt{r^2+a^2}\; \sqrt[4]{a^4+r^4}}.
\end{eqnarray}
The mass is always positive, and the limits are given by $\lim_{r\rightarrow 0}M_\HMS(r)=a/2$  and $\lim_{r\rightarrow \infty}M_\HMS(r)=m$.

If we construct models by varying the (integer) value of $n$, and setting the value of $k$ in~\eqref{newBBS} to zero, we will always have the same qualitative characteristics as the original Simpson--Visser model. Thus, this motivates changing the value of $k$, which we will consider below.

\subsection{Model $n=1$ and $k=2$}\label{SS:n=1+k=2}

For the case $n=1$ and $k=2$ in~\eqref{newBBS}, we have
\begin{eqnarray}
\Sigma(r)=\sqrt{r^2+a^2},\qquad f(r)=1-\frac{2mr^2}{(r^2+a^2)^{3/2}}\label{mod2}\,.
\end{eqnarray}
The function $f(r)$ is identical to that appearing in the regular Bardeen black hole, with the change $a\rightarrow q$. However, the spacetime is completely different to that of Bardeen, due to the term $a^2$ appearing in $\Sigma^2$. (Note that the Bardeen model is recovered by setting  $a\rightarrow 0$ in $\Sigma(r)$, but leaving $a\neq0$ in $f(r)$.) Solving for the roots of $f(r)=0$, we have: i) for $a<a_{ext}=4m/(3\sqrt{3})$, there are four real solutions, which are symmetrical to each other, namely, $(r_+,r_C,-r_C,-r_+)$, (where $r_+$ corresponds to the event horizon and $r_C$ to a Cauchy horizon); ii) for $a=a_{ext}$, we have two real solutions $(r_+,-r_+)$; iii) and for $a>a_{ext}$, no real value exists. 

\begin{figure}[htb!]
	\includegraphics[scale=1]{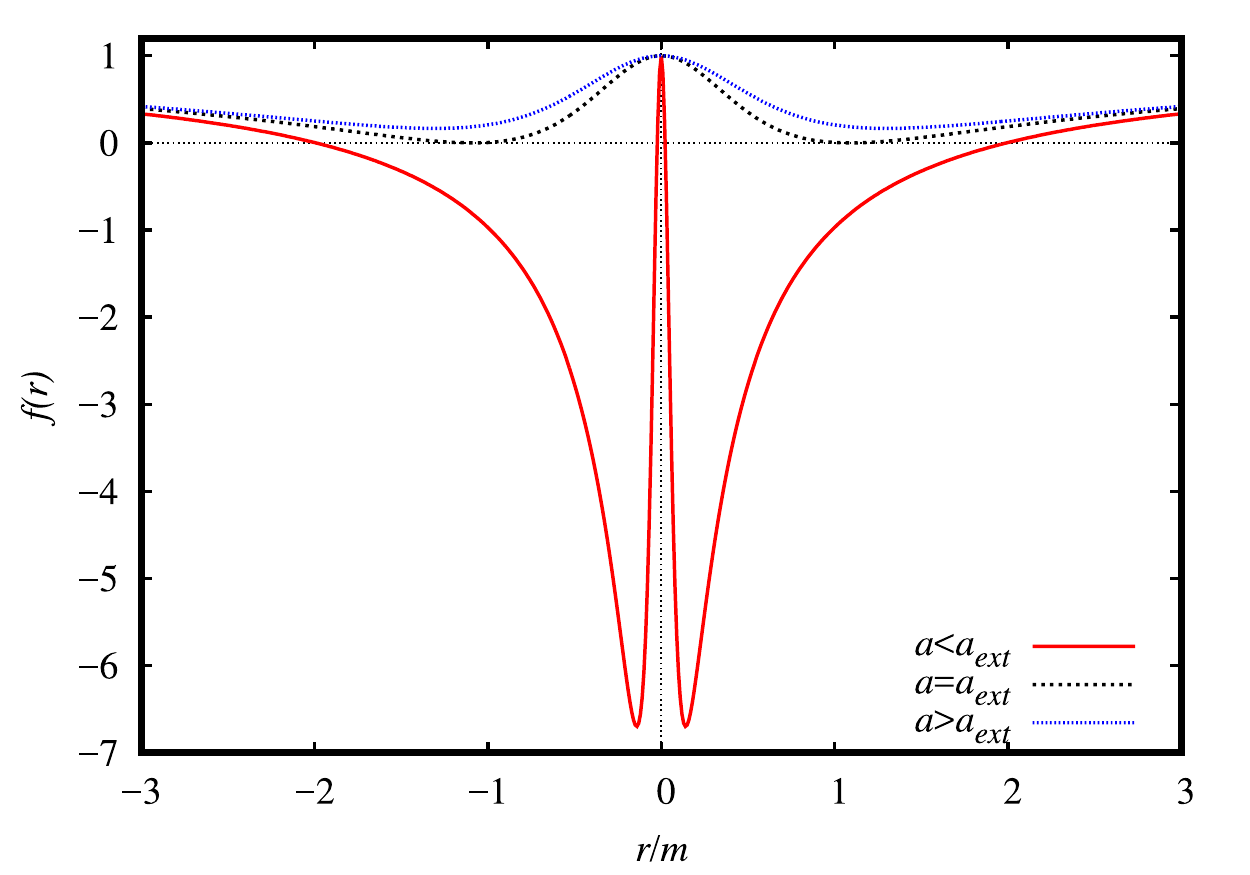}
	\caption{Graphical representation of the possibilities of $f(r)$, given by~\eqref{mod2}. For 
	$a<a_{ext}$, the signature changes four times, which translates as four horizons, two event and two 
	Cauchy horizons. For $a=a_{ext}$, we have a black-bounce with two symmetric degenerate horizons. 
	For $a>a_{ext}$, no horizon exists. See the text for more details.}
	\label{fig1}
\end{figure}

In this new model, we have the first drastic difference compared to the Simpson--Visser model for $a<a_{ext}$, as, by taking the limit $r\rightarrow 0$ in $f(r)$, we verify that $f(r)$ has a positive value with signature $(+,-,-,-)$, contrary to the Simpson--Visser case. This is due to the fact that the latter model only has a single horizon, changing the signature from $(+,-,-,-)$, outside, to $(-,+,-,-)$, inside the horizon, where $r=0$ is contained. However, in the model considered in this section, the signature changes four times, as we can see in Fig.~\ref{fig1} which describes  the behaviour of the metric function $f(r)$. Thus, we have four horizons, namely, two event and two Cauchy horizons.
\enlargethispage{20pt}

The Kretschmann scalar is given by 
\begin{eqnarray}
K=&&\frac{8 m^2 r^4 \left(r^2-2 a^2\right)^2}{\left(r^2+a^2\right)^7}
+\frac{{4}\left(a^2 \left(r^2+a^2\right)^{3/2}+2 m
   r^4\right)^2}{\left(r^2+a^2\right)^7}+\frac{{4} m^2 \left(2 a^4-11 a^2 r^2+2
   r^4\right)^2}{\left(r^2+a^2\right)^7}
   \nonumber\\
   &&\qquad\qquad {
   +\frac{8 \left(a^2 \left(r^2+a^2\right)^{3/2}+m r^2(r^2-4a^2)\right)^2}{\left(r^2+a^2\right)^{{7}}}\,.}
\end{eqnarray}

We verify that for $a>0$ this scalar is finite for all values of $(r,m)$. Thus, the spacetime is always regular. 

\begin{figure}[!htb]
	\includegraphics[scale=0.45]{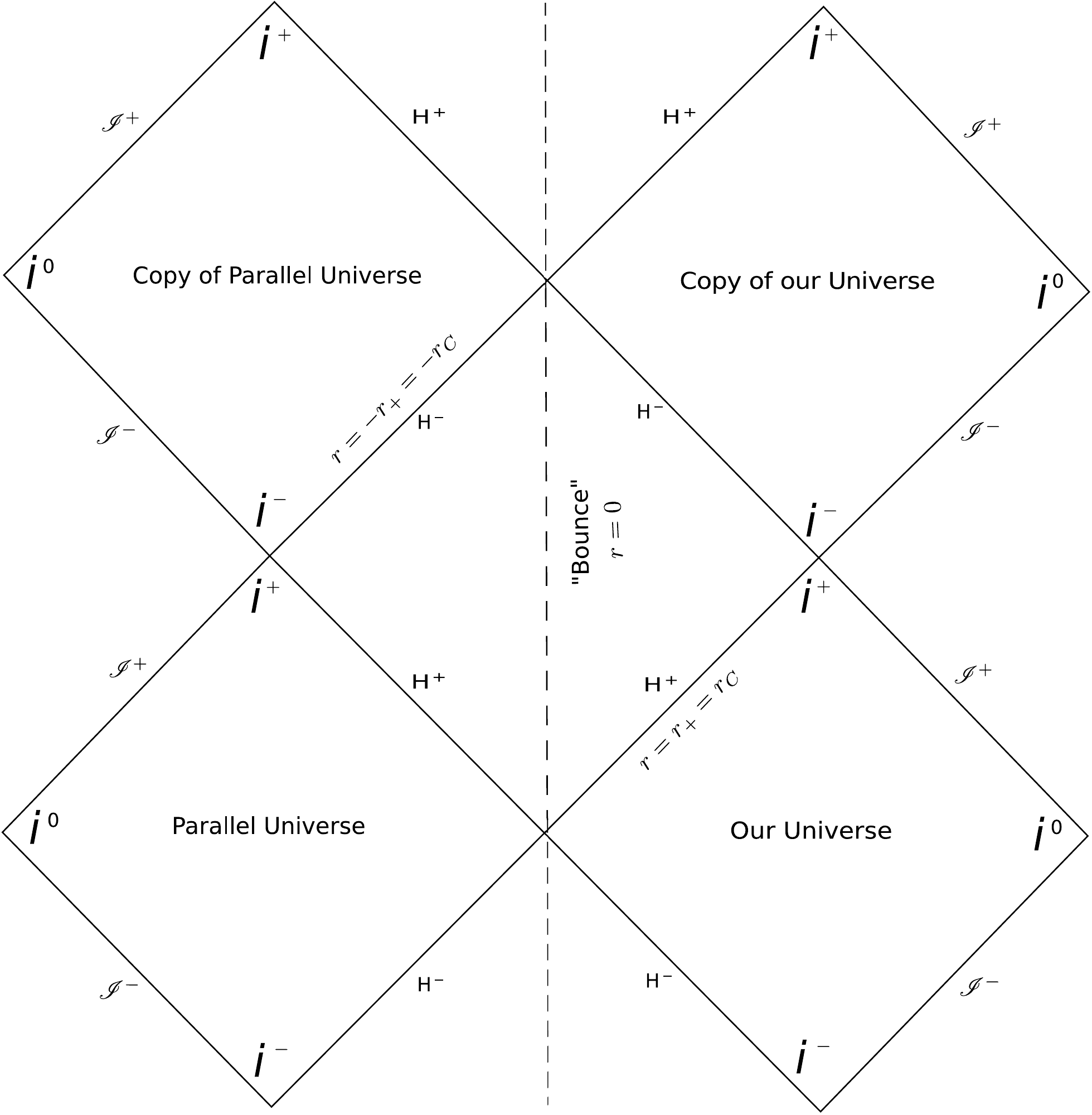}
	\caption{In this example the horizon location is extremal. Mathematically we therefore have repeated roots $r_+ = r_C$ of $f(r) = 0$. Since the extremal horizon ($H_+ = C_+ = r_+$) is as usual an infinite proper distance from any point not on the extremal horizon, the Carter--Penrose diagram is somewhat misleading in that it would be infeasible to propagate through the extremal horizon to reach the hypersurface at $r = 0$. The bounce surface is now timelike, given $f(r)$ does not switch sign through the extremal horizon.}\label{figPenrose4}
\end{figure}

The causal structure is summarized as follows: 
(i) When $a<a_{ext} = \frac{4m}{3\sqrt{3}}$ we have four horizons and  the global causal structure cannot easily be represented in two dimensions, at least not without ``cutting the sheet''. Consequently the usual Penrose diagram approach is not useful. 
(ii) When $a=a_{ext} = \frac{4m}{3\sqrt{3}}$ we have a black-bounce with two symmetric degenerate horizons. The relevant Penrose diagram is depicted in Fig.~\ref{figPenrose4}.
(iii) The specific case of $a>a_{ext} = \frac{4m}{3\sqrt{3}}$ is a horizonless traversable wormhole represented by the Penrose diagram in Fig.~\ref{figPenrose1}.

In the region where the $t$ coordinate is timelike, the energy conditions for this model are given by the following relations:
\begin{eqnarray}
&&
NEC_1\Longleftrightarrow
  -\frac{2 a^2 \left(1-\frac{2 m r^2}{\left(r^2+a^2\right)^{3/2}}\right)}
{\kappa ^2 \left(r^2+a^2\right)^2}\geq0\,, 
\qquad\qquad\;
NEC_2\Longleftrightarrow
    \frac{ma^2(13 r^2-2 a^2)}{\kappa ^2 \left(r^2+a^2\right)^{7/2}}\geq0 \,,\,  
\\
&&
WEC_3\Longleftrightarrow
   -\frac{a^2 \left(\left(r^2+a^2\right)^{3/2}-8 m r^2\right)}{\kappa ^2 \left(r^2+a^2\right)^{7/2}}\geq0 \,,
   \qquad
SEC_3\Longleftrightarrow
  \frac{2 m a^2 \left(7r^2-2 a^2\right)}
{\kappa ^2 \left(r^2+a^2\right)^{7/2}}\geq0 \,,
\\
&&
DEC_1\Longrightarrow  \frac{12 a^2 m r^2}{\kappa ^2 \left(r^2+a^2\right)^{7/2}}
\geq0 \,, \qquad \,
DEC_2\Longrightarrow - \frac{a^2 \left( 2 (a^2+r^2)^{3/2} - m \left(2 a^2+3 r^2\right)\right) }{\kappa ^2 \left(r^2+a^2\right)^{7/2}}  \,.
\end{eqnarray}

As mentioned above, the $NEC_1=WEC_1=SEC_1$ are violated for $|r|>r_+$; the $DEC_{2}$ and $WEC_3$ are violated for $|r|\gg r_+$. The $NEC_2$ is violated in the range $-\sqrt{2/13}\;a<r<\sqrt{2/13}\;a$, and finally the $SEC_3$ is violated for $-\sqrt{2/7}\;a<r<\sqrt{2/7}\;a$. As $WEC_3=\rho$ is violated outside the event horizon, we continue to have negative energy densities.

In the region where the $t$ coordinate is spacelike, the energy conditions for this model are given by the following relations:
\begin{eqnarray}
&&
NEC_1\Longleftrightarrow
  \frac{2 a^2 \left(1-\frac{2 m r^2}{\left(r^2+a^2\right)^{3/2}}\right)}
{\kappa ^2 \left(r^2+a^2\right)^2}\geq0\,, \qquad\qquad\;
NEC_2\Longleftrightarrow
    \frac{2 a^2(r^2+a^2)^{3/2} +a^2m(9r^2-2a^2)}{\kappa ^2 \left(r^2+a^2\right)^{7/2}}\geq0 \,,\,   
\\
&&
WEC_3\Longleftrightarrow
   \frac{a^2 \left(\left(r^2+a^2\right)^{3/2}+4 m r^2\right)}{\kappa ^2 \left(r^2+a^2\right)^{7/2}}\geq0 \,,\qquad
SEC_3\Longleftrightarrow
  \frac{4a^2(r^2+a^2)^{3/2} +2 m a^2 \left(3r^2-2a^2\right)}
{\kappa ^2 \left(r^2+a^2\right)^{7/2}}\geq0 \,,
\\
&&
DEC_1\Longrightarrow  \frac{12 a^2 m r^2}{\kappa ^2 \left(r^2+a^2\right)^{7/2}}\geq0 \,, \qquad \,
\qquad\qquad\quad
DEC_2\Longrightarrow  -\frac{a^2 m \left(r^2-2a^2\right)}{\kappa ^2 \left(r^2+a^2\right)^{7/2}}
\geq0  \,.
\end{eqnarray}
Again, the $NEC_1=WEC_1=SEC_1$ is violated for sub-horizon regions.

The Hernandez--Misner--Sharp mass~\eqref{mass} for this model~\eqref{mod2} is given by
\begin{eqnarray}
M_\HMS(r)=\frac{a^2}{2\sqrt{r^2+a^2}}+\frac{mr^4}{\left(r^2+a^2\right)^2}.
\end{eqnarray}
The mass is always positive and possesses the  limits $\lim_{r\rightarrow 0}M_\HMS(r)=a/2$  and $\lim_{r\rightarrow \infty}M_\HMS(r)=m$.

If one wishes to construct models by setting $k = 2 $ and by changing the integer $n$, we verify that the respective spacetime possesses similar characteristics as the case considered above, and the energy density will always be negative for the region outside the event horizons.

\subsection{Model with zero energy density}\label{SS:zero}
\def\O{{\mathcal{O}}}

While (as we have seen above) some of the energy conditions will always be violated, can we at least satisfy the $WEC_3$? This would require a non-negative energy density, and we shall first consider the special case where the energy density is identically zero.  Setting $\rho(r)=0$ and $\Sigma(r)=\sqrt{r^2+a^2}$ in \eqref{density+} and solving the differential equation for $f(r)$, we obtain
\begin{equation}
f(r) = {(\sqrt{r^2+a^2}+K)\;\sqrt{r^2+a^2}\over r^2}.
\end{equation}
But the regularity condition that $f(0)$ be finite requires the integration constant $K$ to be set to $K=-a$, in which case
\begin{equation}
f(r) = {(\sqrt{r^2+a^2}-a)\;\sqrt{r^2+a^2}\over r^2}.
\end{equation}
This geometry is horizonless and
\begin{equation}
f(r) = {1\over2} + \O(r^2) ; 
\qquad\qquad  \hbox{while} \qquad\qquad
f(r) = 1 -{a\over r} + \O(1/r^2).
\end{equation}
From the Einstein equations applied to this metric it is now easy to verify that $\rho=0$ and that
\begin{equation}
p_r = -{2a^2(\sqrt{r^2+a^2}-a)\over r^2(r^2+a^2)^{3/2}}; \qquad\qquad
p_t = {2\sqrt{r^2+a^2}(3a^2+2 r^2) - a(7r^2 +6a^2) \over 2(r^2+a^2)^{3/2} r^4}.
\end{equation}
For this model $p_r<0$ for any nonzero $r$, and $p_t>0$ for any $r$, so certainly $NEC_1$ is still violated throughout the spacetime. $WEC_3$ is by construction marginally satisfied. So while we can force the WEC to be tolerably well-behaved, other energy conditions will still be violated.
The Hernandez--Misner--Sharp mass for this spacetime is particularly simple, $M_\HMS(r) = a/2$ everywhere. 

One could try to generalize this construction by choosing some positive function $\rho_*(r)>0$ and setting
$\rho(r) = \rho_*(r)>0$. One would then solve the differential equation for $f(r)$ arising from \eqref{density+}, fixing the integration constant by demanding the finiteness of $f(0)$.
Such a construction would at least satisfy $WEC_3$, at least in the domain of outer communication, 
but the status of the other energy conditions would remain unresolved.

However when it comes to analyzing the $NEC_1=WEC_1=SEC_1$ we can say more. From equation \eqref{cond1b} substituting $\Sigma\to \sqrt{r^2+a^2}$ , we can see that:
\begin{equation}
NEC_{1}=WEC_1=SEC_1 \Longleftrightarrow
-\frac{2 |f(r)| a^2}{\kappa ^2 (r^2+a^2)}\geq 0,\label{cond1c}\\
\end{equation}
So $NEC_1=WEC_1=SEC_1$ are definitely violated everywhere except on the horizons themselves, indeed whenever $f(r)\neq 0$.

\subsection{Model $M(r)=m\cos^{2n}\left[r_0/\Sigma(r)\right]$}\label{SS:M(r)_1}

Choosing $M(r)=m\cos^{2n}\left[r_0/\Sigma(r)\right]$,  we have
\begin{eqnarray}
f(r)=1-\frac{2M(r)}{\Sigma(r)}=1-\frac{2m\cos^{2n}\left[r_0/\Sigma\right]}{\Sigma}\label{mod3}\,,
\end{eqnarray}
where for $n\rightarrow 0$, since we are retaining $\Sigma= \sqrt{r^2+a^2}$, we recover the Simpson--Visser model. As the radial coordinate tends to infinity we have $\lim_{r\rightarrow \infty}f(r)=1$. (Indeed, for $r\gg1$ we have $f(r)\sim 1-(2m/r)$.) Also
\begin{eqnarray}
\lim_{r\rightarrow 0}f(r)=1-\frac{2m\cos^{2n}\left(r_0/a\right)}{a}\,.
\end{eqnarray}
Appropriately choosing $r_0$ we have three possibilities, namely: $f(0)>0$, with the signature $(+,-,-,-)$; $f(0)=0$ with $2m\geq a$; and $f(0)<0$, with signature $(-,+,-,-)$.

\begin{figure}[htb!]
	\includegraphics[scale=.6]{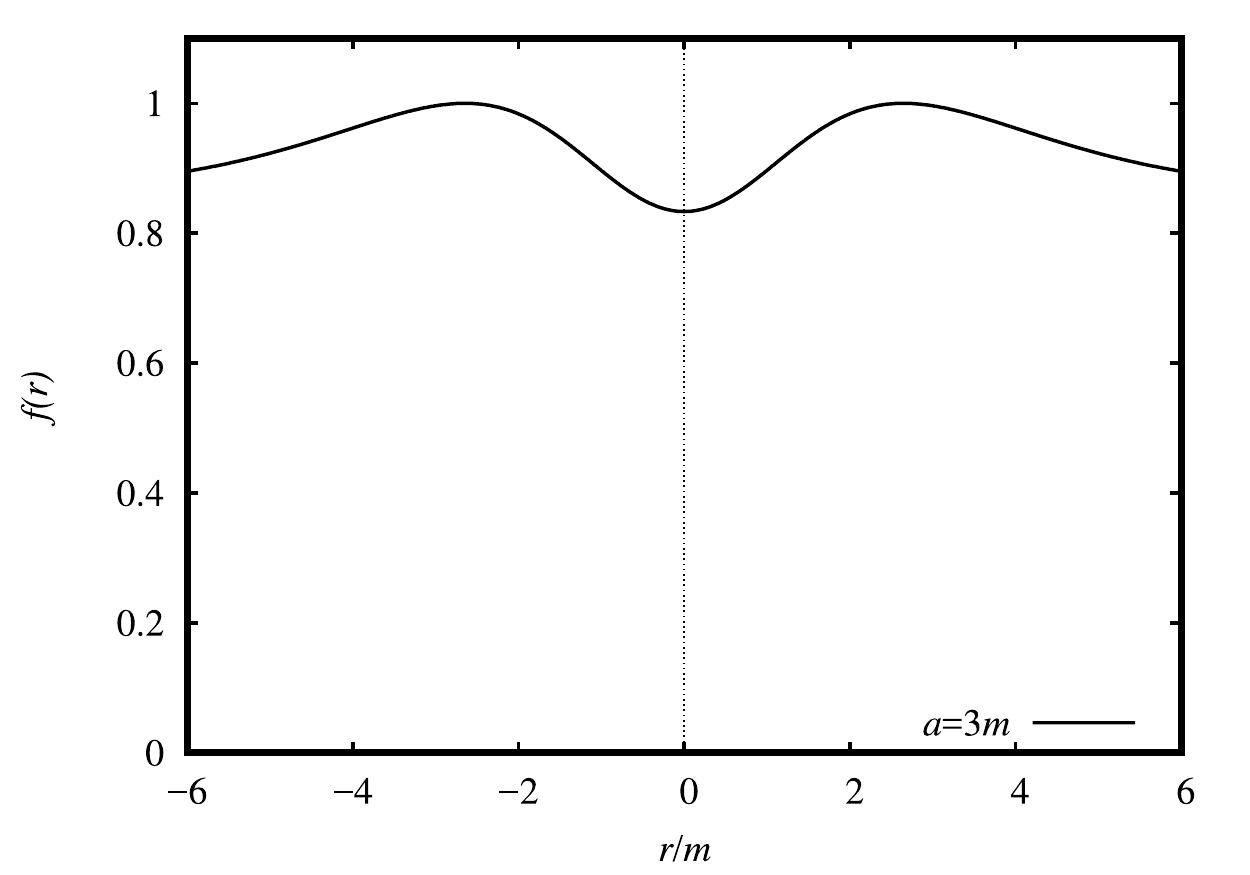}
	\includegraphics[scale=.6]{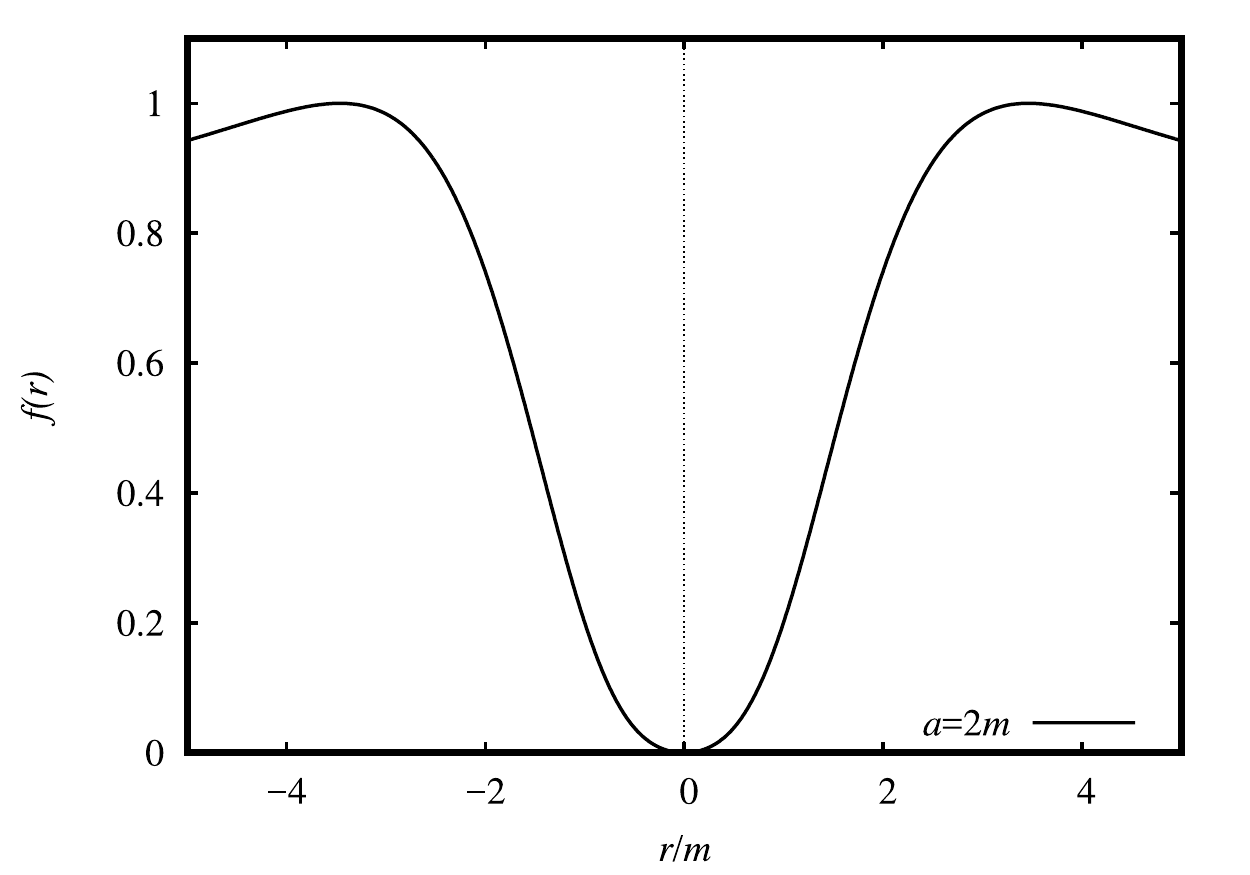}
	\includegraphics[scale=.6]{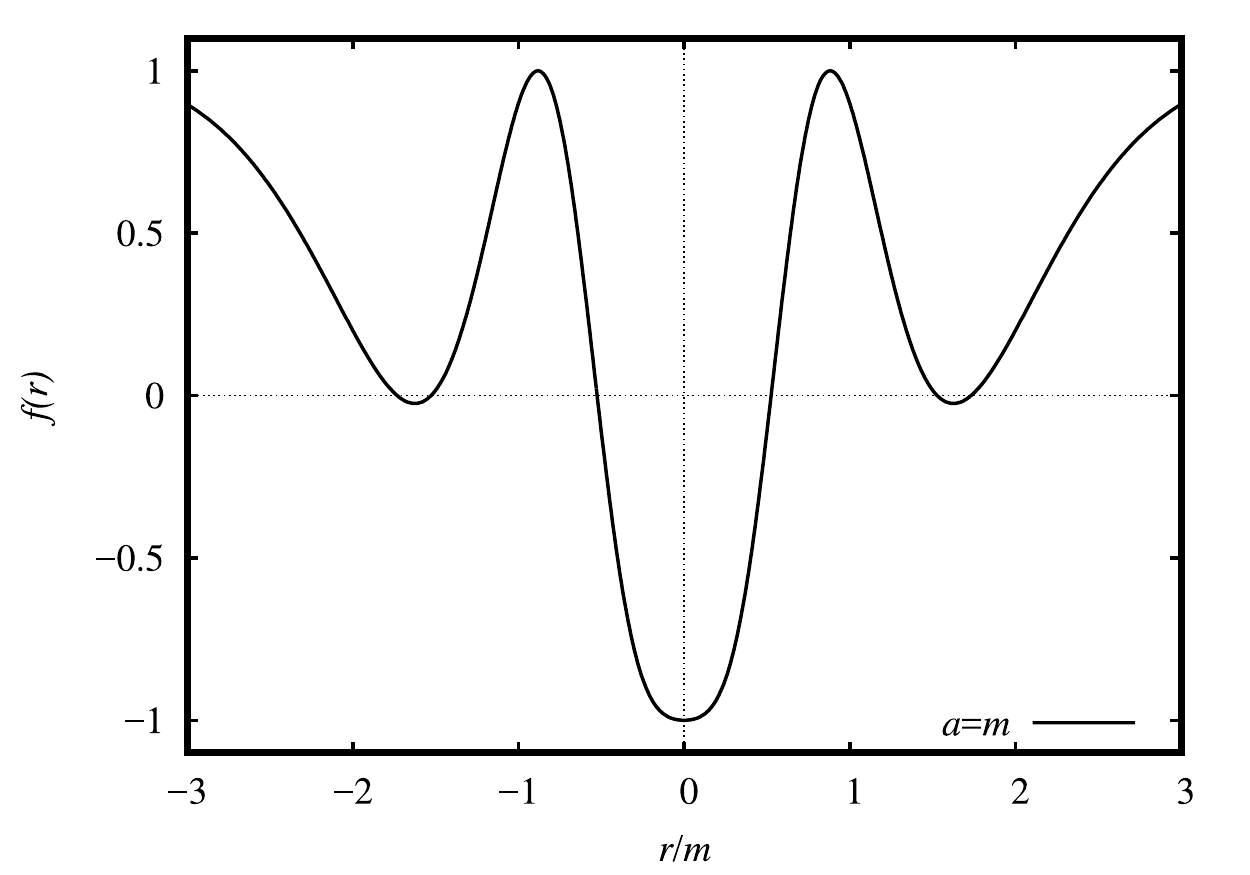}
	\includegraphics[scale=.6]{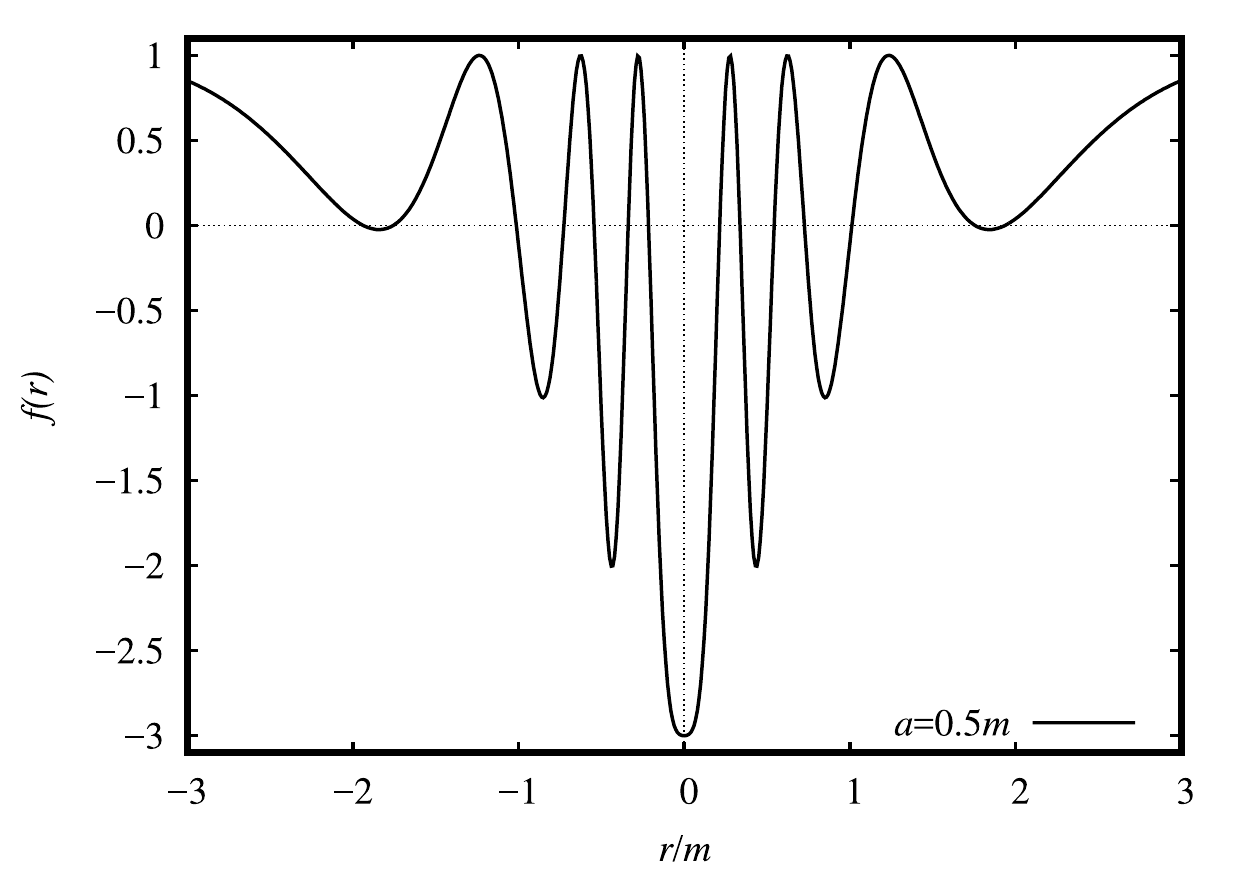}
	\caption{Graphical representation of the possibilities of $f(r)$, given by~\eqref{mod3} with $r_0=2\pi a$.}\label{fig3}
\end{figure}    

\begin{figure}[htb!]
	\includegraphics[scale=.6]{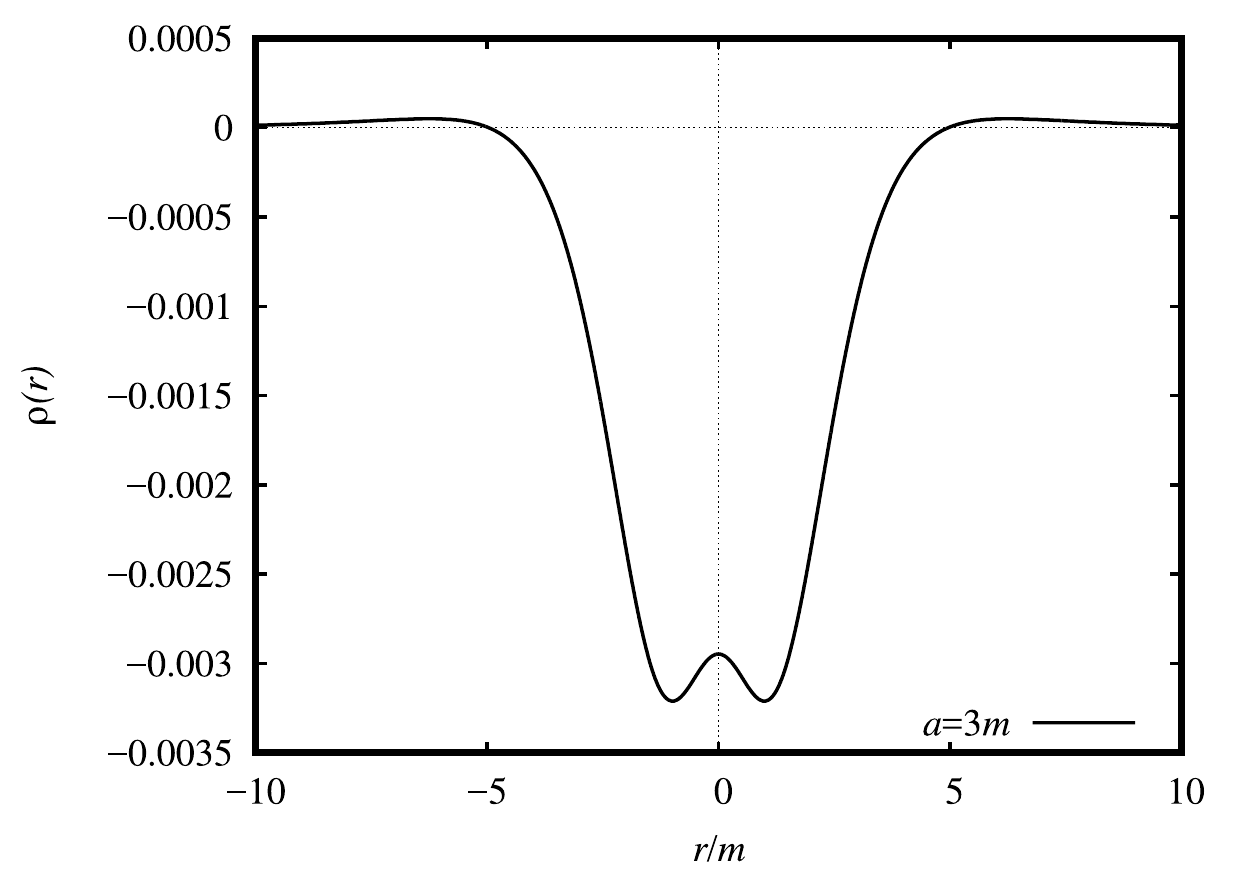}
	\includegraphics[scale=.6]{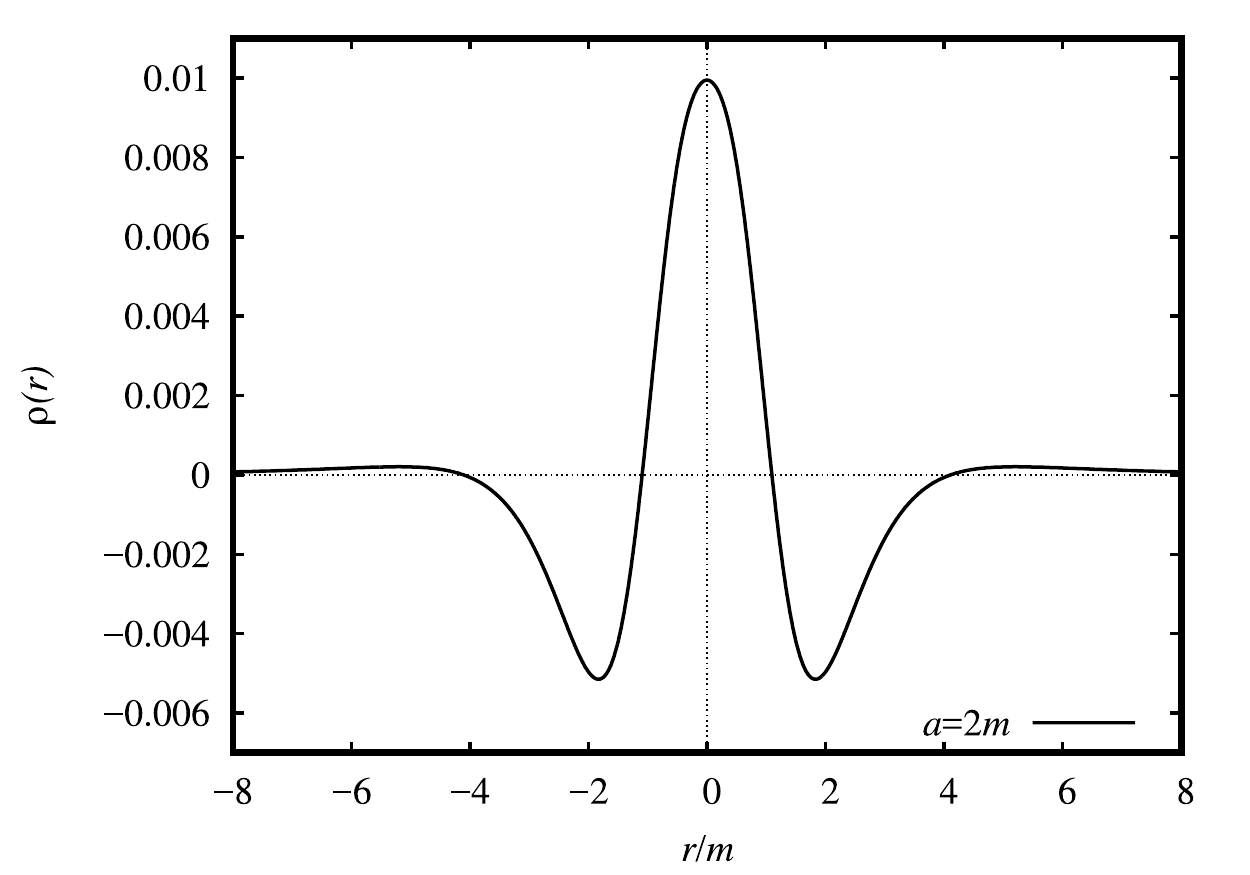}
	\caption{Graphical representation of the energy density $\rho(r)$, to the model~\eqref{mod3} with $r_0=2\pi a$.}\label{fig4}
\end{figure} 

The number of horizons may also be modified by changing $f(0)$, \emph{i.e.}, $m/a$, as we see in Fig.~\ref{fig3}. In the plots of Fig.~\ref{fig3}, one may envision the three structures of a black-bounce: (i) for $r_0=2\pi a$ and $a=3m$, we have no horizons, and consequently we have a wormhole with a two-way timelike throat at $r=0$; (ii) for $r_0=2\pi a$ and $a=2m$, there is an extremal throat at $r=0$; (iii) for the specific example $r_0=2\pi a$ and $a=0.5m$, we have a regular black hole with $14$ horizons, and where $r=0$ is a bounce. Thus, the number of horizons can in principle grow indefinitely. For the latter case, the causal structure cannot be represented in a Penrose diagram. The causal structure is given by: (i) when $a>a_{ext} =2m$, $(r_0 =2\pi a)$, in Fig.~\ref{figPenrose1}; (ii) when $a=a_{ext} =2m$, $(r_0 =2\pi a)$, in Fig.~\ref{figPenrose2}.

Analytically, the energy density is not particularly simple; we represent it in Fig.~\ref{fig4}. We see that the energy density oscillates as the function $f(r)$. Asymptotically expanding the energy density as $r\rightarrow \infty$ the dominant term is 
\begin{equation}
\rho(r) \approx -a^2/\kappa^2 r^4.
\end{equation} 
Therefore  the energy density is certainly negative for some regions outside the event horizon.

The Hernandez--Misner--Sharp mass~\eqref{mass} for the model~\eqref{mod3} is given by
\begin{eqnarray}
M_\HMS(r)=\frac{a^2}{2\sqrt{r^2+a^2}}+\frac{mr^2 \cos^{2n}\left(r_0/\sqrt{r^2+a^2}\right)}{r^2+a^2}.
\end{eqnarray}
The mass is always positive with the limits $\lim_{r\rightarrow \infty}M_\HMS(r)=m$ and $\lim_{r\rightarrow 0}M_\HMS(r)=a/2$.

\subsection{Model $M(r)=m\arctan^n(r/a)\;(\Sigma/r) (2/\pi)^n$}\label{SS:M(r)_2}

\begin{figure}[htb!]
	\includegraphics[scale=0.6]{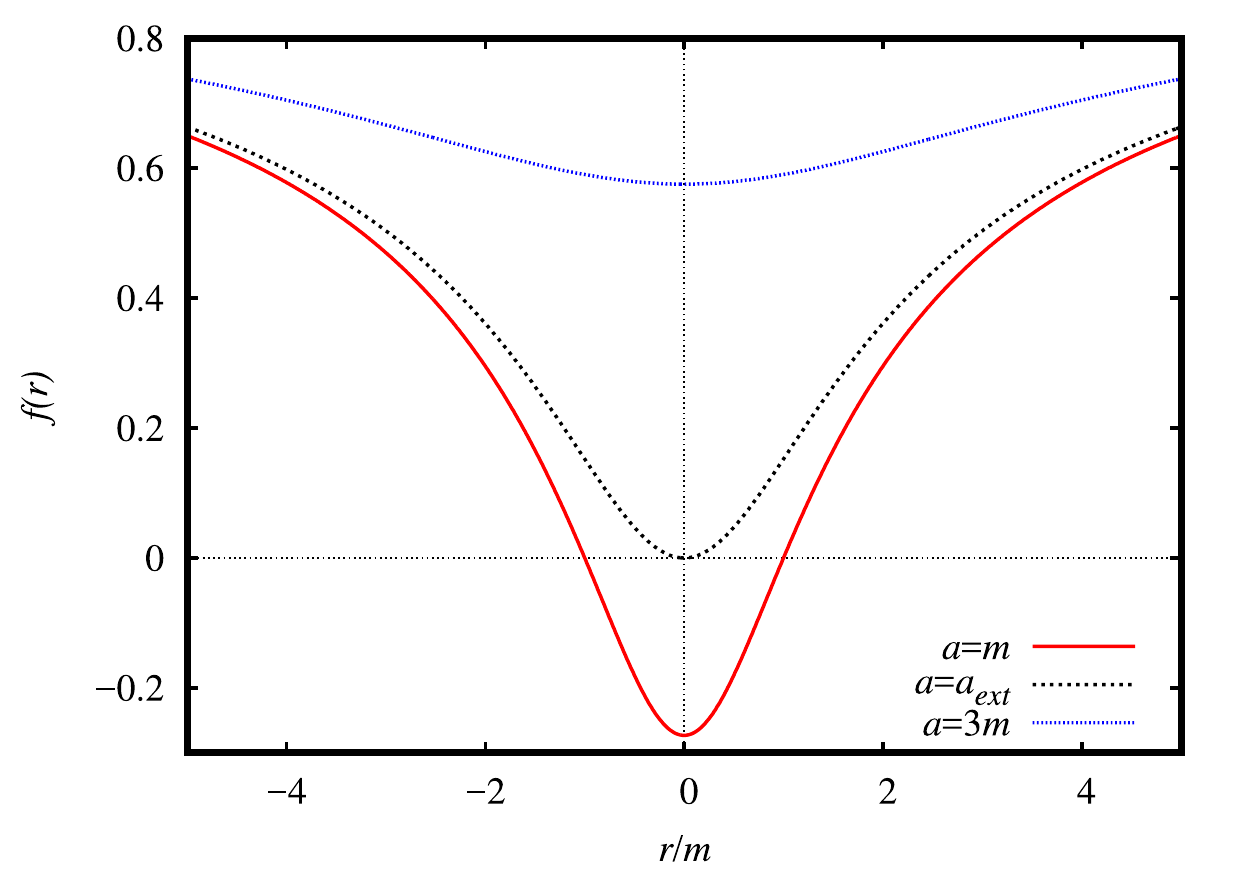}\qquad \includegraphics[scale=0.6]{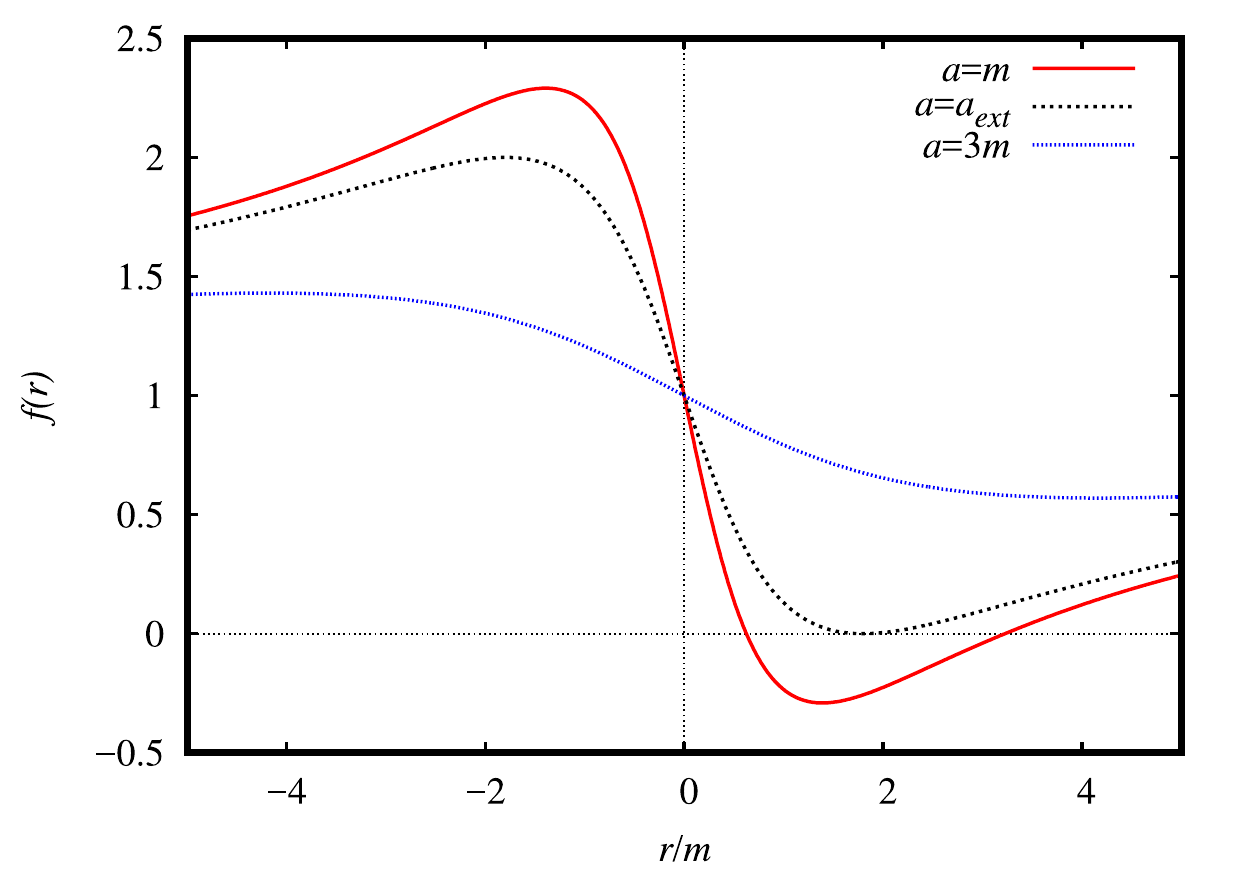}
	\caption{Graphical representation of $f(r)$ for~\eqref{mod4}. In the left side $n=1$ and right side $n=2$. For $n=2$, the extreme value is approximately $a_{ext}\approx5.16315560586775m/\pi^2$.}\label{fig5}
\end{figure}

We will now define a mass function that provides a positive energy density. More specifically, consider the specific case $M(r)=m\arctan^n(r/a)(\Sigma/r)(2/\pi)^n $, so that the metric function $f(r)$ is given by
\begin{eqnarray}
f(r)=1-\frac{2M(r)}{\Sigma(r)}=1-\frac{2m\arctan^n(r/a)}{r}\left(\frac{2}{\pi}\right)^n\label{mod4}\,.
\end{eqnarray}
In the limit $(a,n)\rightarrow 0$ we regain the Schwarzschild solution. One may show that the Kretschmann scalar is regular, so the spacetime is regular. We can now fix $n$ and regulate the presence of horizons by adjusting $a$, as shown in Fig.~\ref{fig5}. For instance, consider $n=1,2$, where the extreme case for $n=1$ is given by $a_{ext}=4m/\pi$, and for $n=2$ we have $a_{ext}\approx 5.16315560586775m/\pi^2$.

\begin{figure}[htb!]
	\includegraphics[scale=0.5]{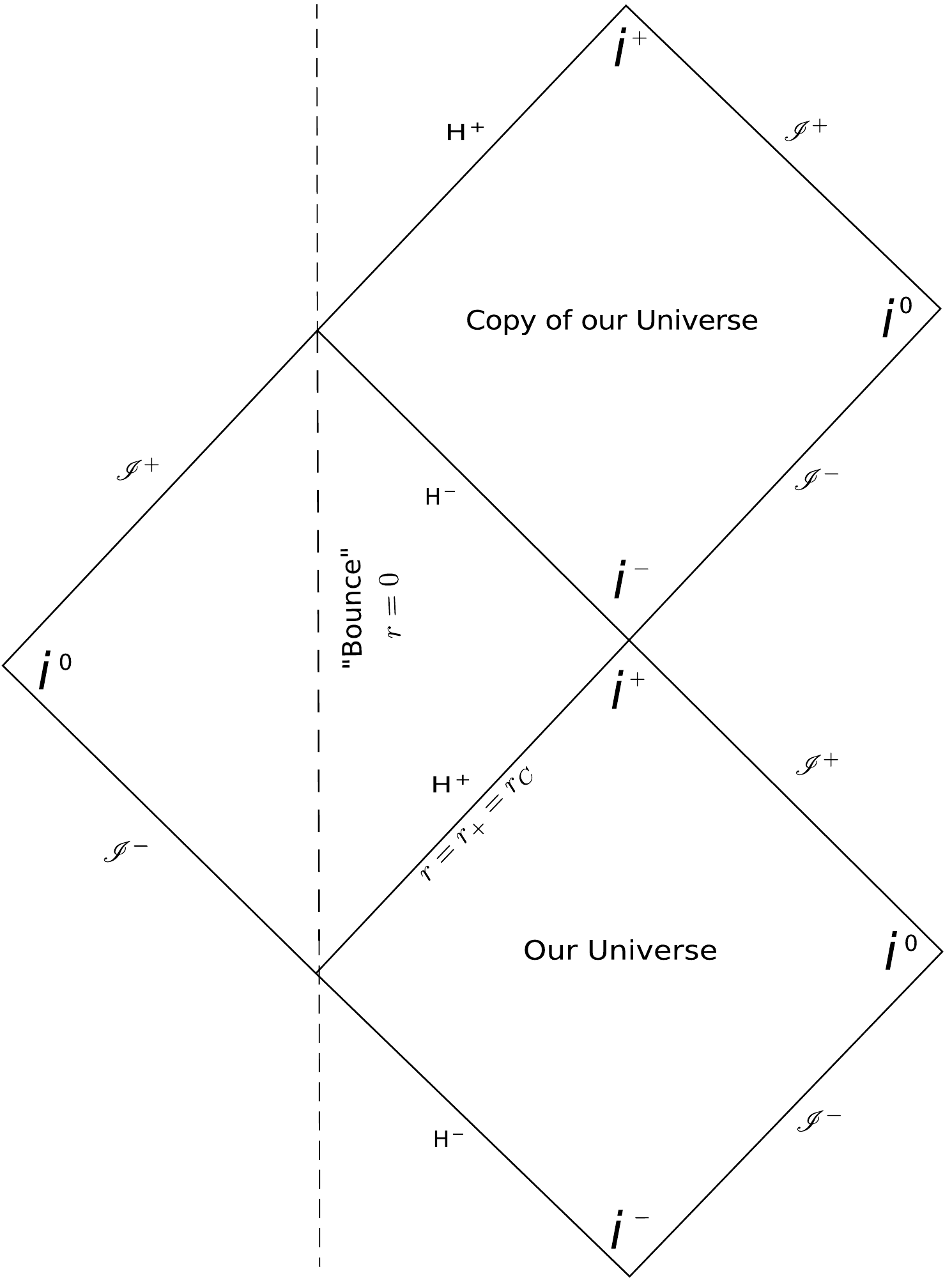}
	\caption{Carter--Penrose diagram for the case where we have an extremal horizon in our universe, followed by a timelike ``bounce'' hypersurface at $r=0$, bouncing into a separate universe without horizons. Since the extremal horizon ($H+ = C+ = r+)$ is as usual an infinite proper distance from any point not on the extremal horizon, the Carter--Penrose diagram is somewhat misleading in that it would be infeasible to propagate through the extremal horizon to reach the hypersurface at $r=0$.\\
	}
	\label{figPenrose5}
\end{figure}

\begin{figure}[htb!]
	\includegraphics[scale=0.7]{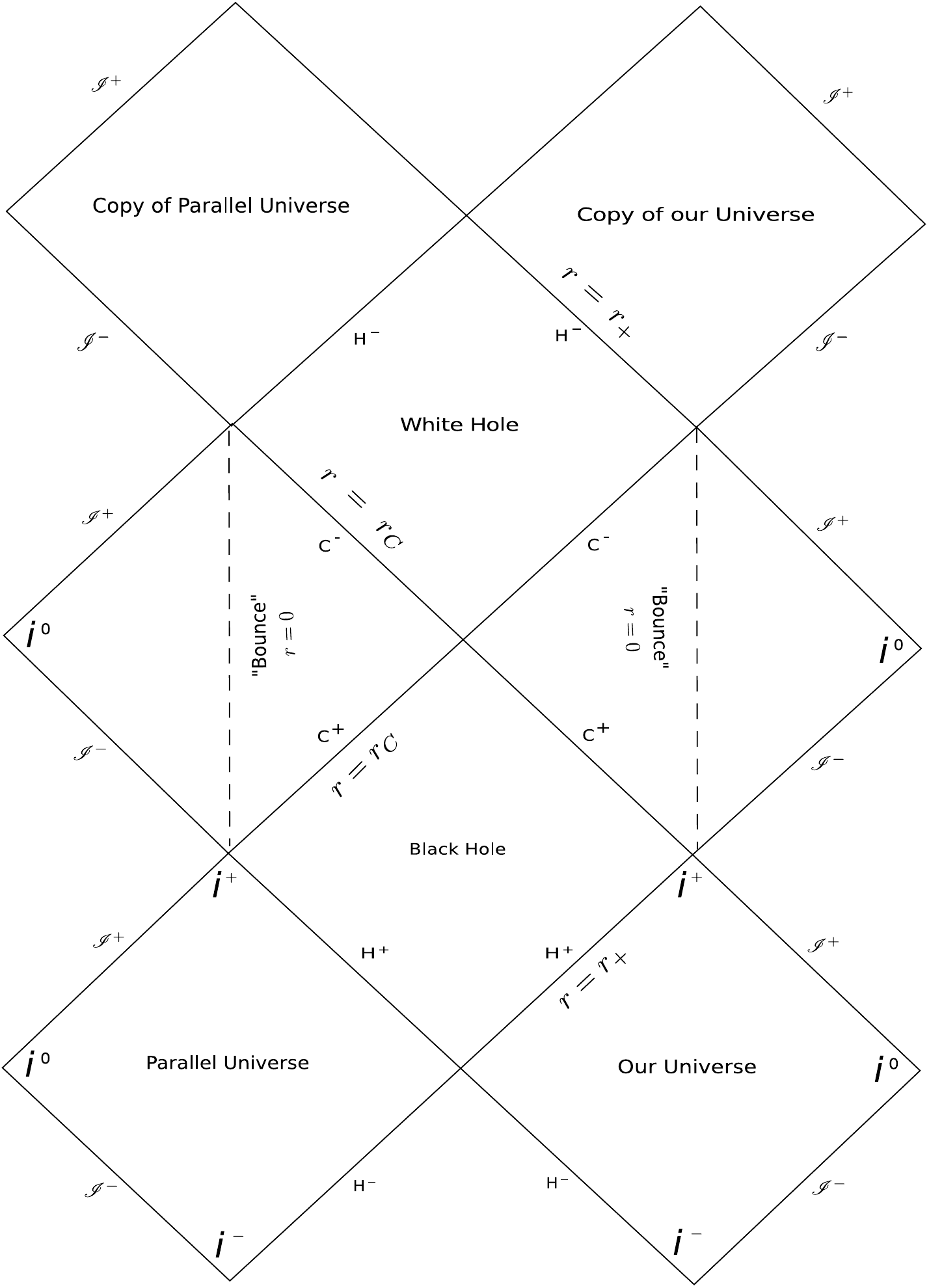}
	\caption{Carter--Penrose diagram for the case where we have an inner and outer horizon in our universe, followed by a timelike ``bounce'' hypersurface at	$r=0$, bouncing into a separate universe without horizons.\\
	}
	\label{figPenrose6}
\end{figure}

The causal structure is given by the Penrose diagrams, namely:
(i) the cases $n=1$ and $a>a_{ext} =4m/\pi $; $n=2$ and $a>a_{ext} \approx 5.16m$, are depicted in Fig.~\ref{figPenrose1};
(ii) for $n=1$, $a=a_{ext} =4m/\pi$, in Fig.~\ref{figPenrose2};
(iii) for $n=1$, $a<a_{ext} =4m/\pi$, in  Fig.~\ref{figPenrose3};
(iv) for $n=2$, $a=a_{ext} \approx 5.16m$, is depicted in Fig.~\ref{figPenrose5}, where we have an extremal horizon in our universe, followed by a timelike ``bounce'' hypersurface at $r = 0$, bouncing into a separate universe without horizons;
(v) for $n=2$ and $a=a_{ext} \approx 5.16m$, in  Fig.~\ref{figPenrose6}, where here we have an inner and outer horizon in our universe, followed by a timelike ``bounce'' hypersurface at $r = 0$, bouncing into a separate universe without horizons.

\begin{figure}[htb!]
	\includegraphics[scale=0.6]{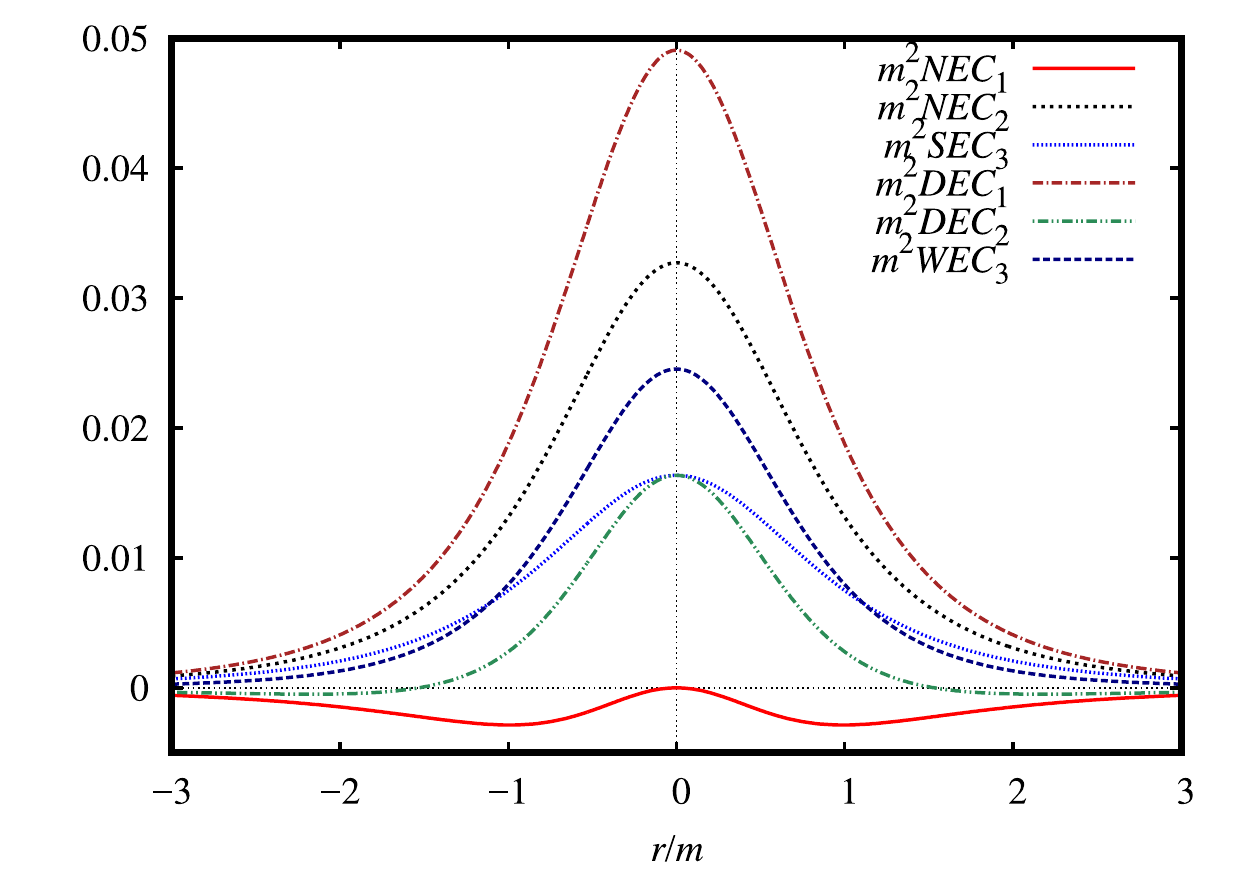}\qquad \includegraphics[scale=0.6]{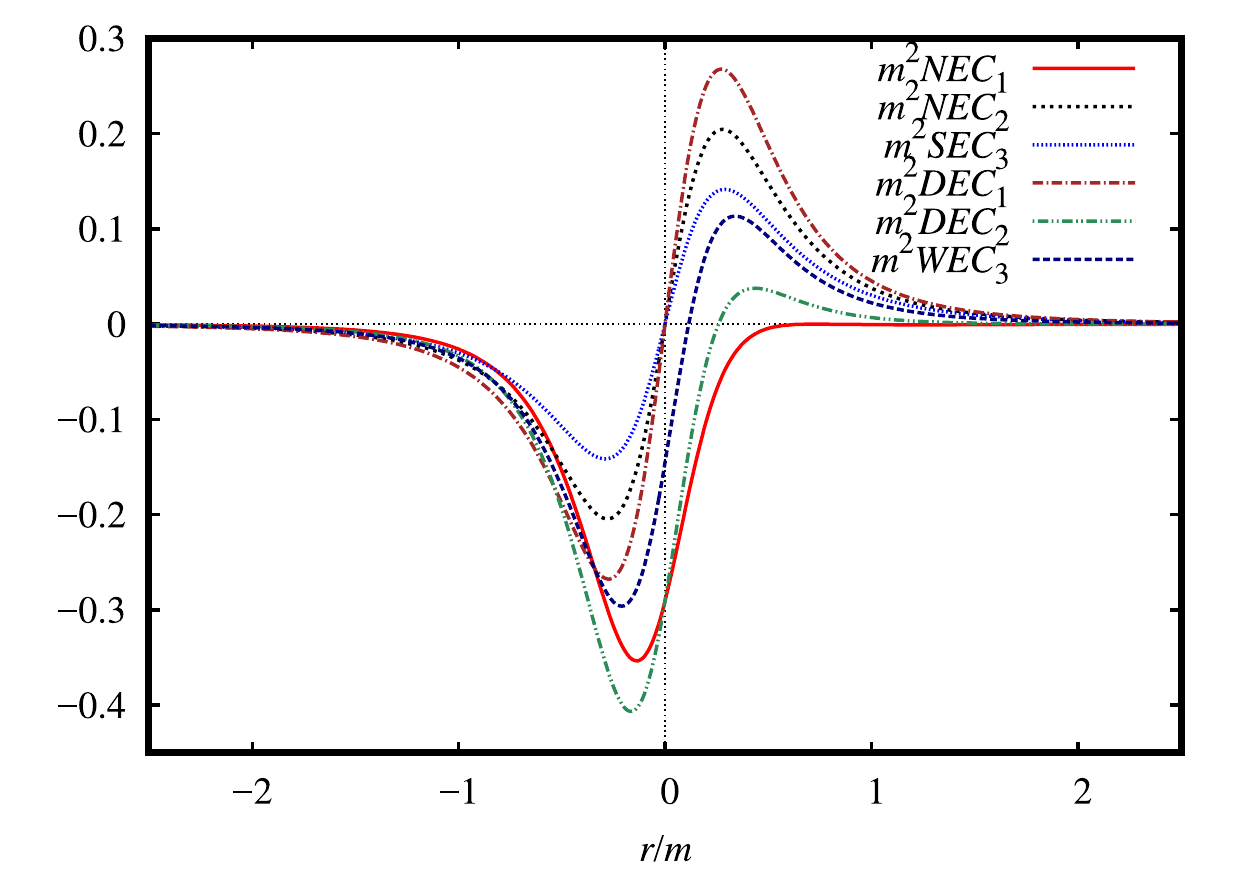}
	\caption{Graphical representation of the energy conditions for~\eqref{mod4} with $n=1, a=a_{ext}=4m/\pi,\kappa^2=8\pi$, in the left plot, and $n=2, a=a_{ext} \approx 5.16m,\kappa^2=8\pi$ in the right plot.}\label{fig6}
\end{figure}

The energy conditions are shown in Fig.~\ref{fig6}. In both cases, one verifies that all the energy conditions are violated, but in particular, the condition $WEC_3=\rho$ is always satisfied for the cases $n=1$, $a<a_{ext}$ and $a=a_{ext}$; this renders a positive energy density everywhere. A new feature that highlights the difference between a regular black-bounce solution (with horizons) and the standard regular black hole, considered in the Introduction, is that the condition $SEC_3$, defined in~\eqref{Econd1}, is always satisfied, as one can readily verify in Fig.~\ref{fig6}. More specifically, this does not occur for regular spherically symmetrical black holes, where this condition is always violated within the event horizon~\cite{Zaslavskii}.

The Hernandez--Misner--Sharp mass~\eqref{mass} for the model~\eqref{mod4} is given by
\begin{eqnarray}
M_\HMS(r)=\frac{\pi ^{-n} \left(a^2 \pi ^n+m 2^{n+1} r \arctan\left(\frac{r}{a}\right)^n\right)}{2 \sqrt{r^2+a^2}}\label{massmod4}\,.
\end{eqnarray}
Note that for $n$ odd the mass is always positive, and we have the following limits: $\lim_{r\rightarrow \infty}M_\HMS(r)=m$; and $\lim_{r\rightarrow 0}M_\HMS(r)=a/2$. 
%

\subsection{Model $M(r)=m\arctan^n(r/a)(2/\pi)^n$}\label{SS:M(r)_3}

\begin{figure}[htb!]
	\includegraphics[scale=0.6]{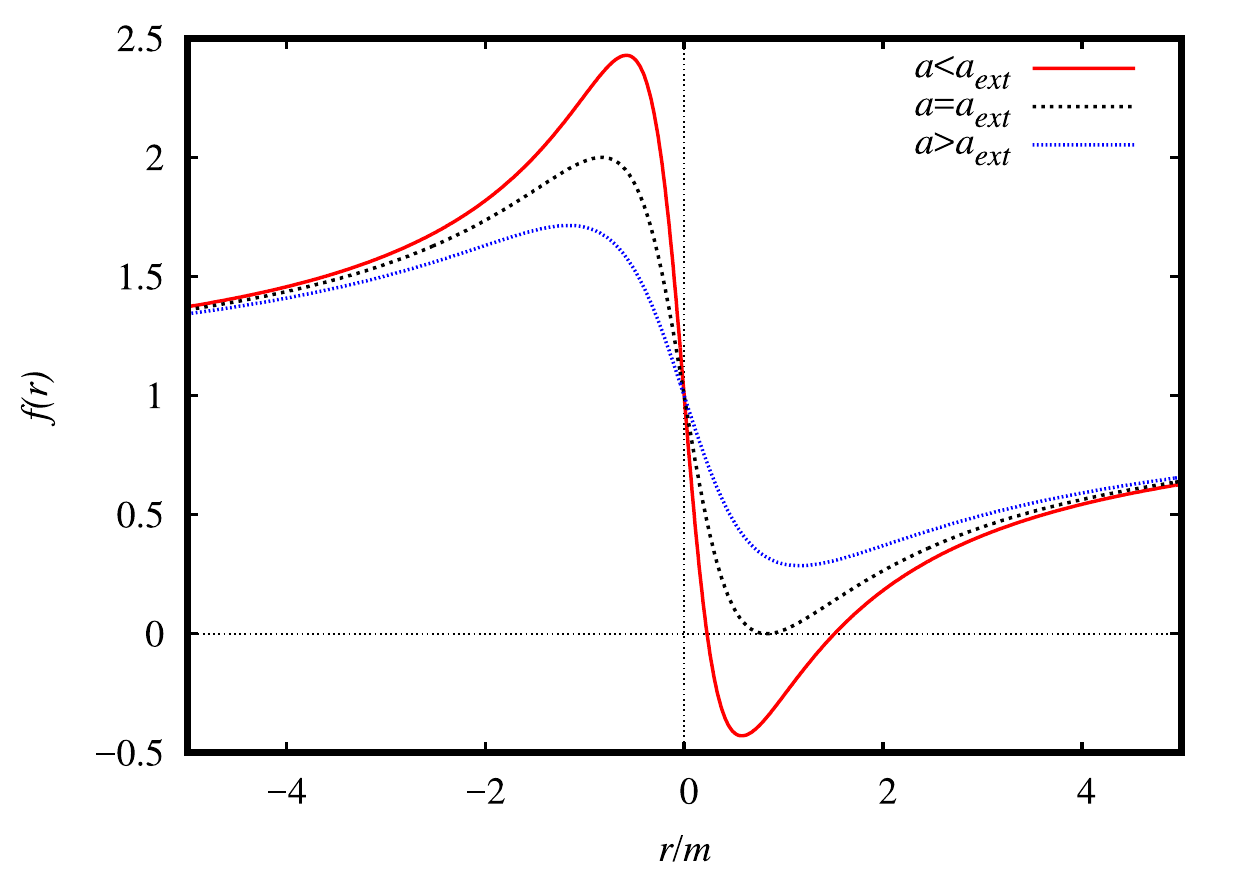}\qquad
	\includegraphics[scale=0.6]{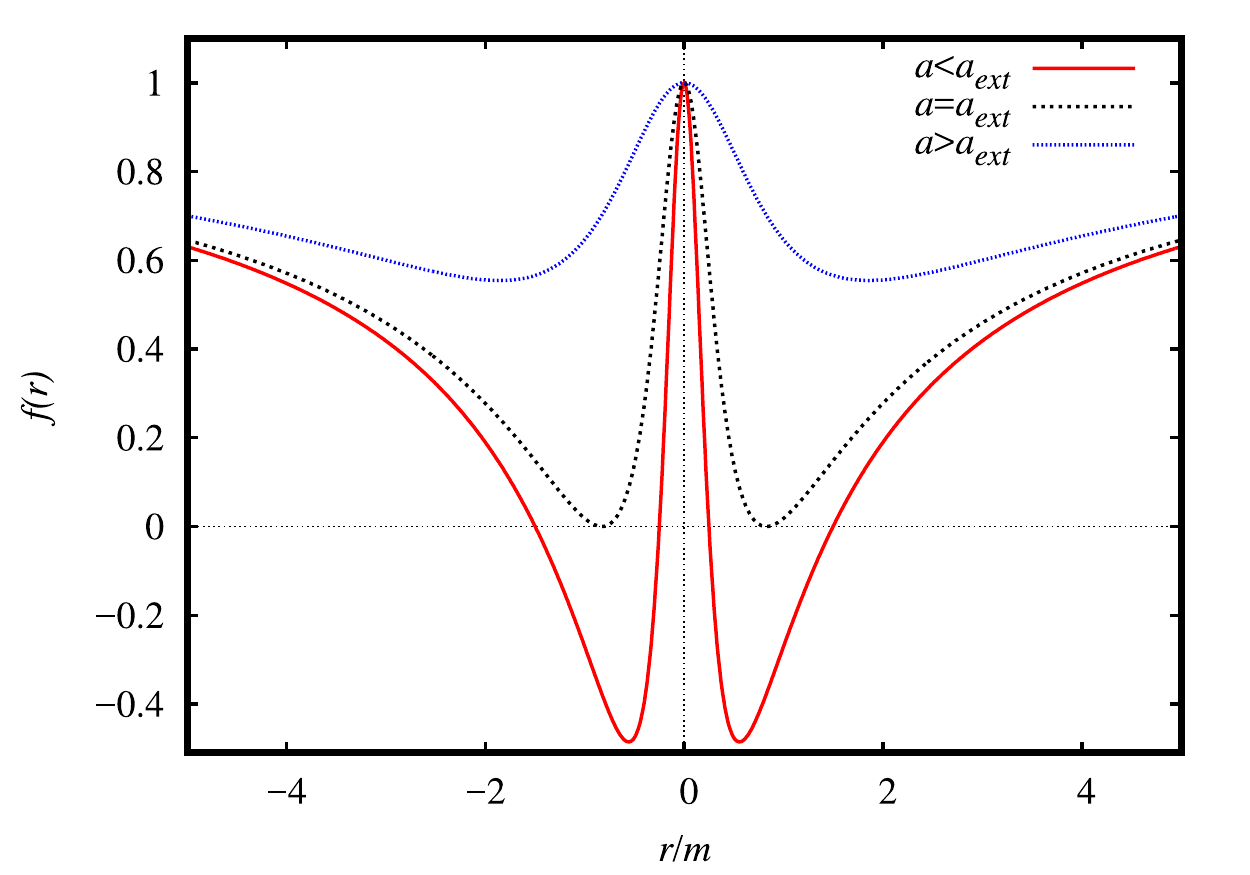}
	\caption{Graphical representation of $f(r)$ for~\eqref{mod5}, with $n=1$ (left) and $n=2$ (right) for different values of $a$.}\label{fig9}
\end{figure}

We now define a mass function that provides a positive energy density, given by $M(r)=m\arctan^n(r/a)(2/\pi)^n$, so that the factor $f(r)$ takes the form
\begin{eqnarray}
f(r)=1-\frac{2M(r)}{\Sigma(r)}=1-\frac{2m\arctan^n(r/a)}{\sqrt{r^2+a^2}}\left(\frac{2}{\pi}\right)^n\label{mod5}\,.
\end{eqnarray}
For $n\rightarrow 0$ we have the Simpson--Visser spacetime and $(a,n)\rightarrow 0$ the Schwarzschild solution. The Kretschmann scalar is regular everywhere. Fig.~\ref{fig9} shows $f(r)$, where for the cases $n=1$ and $n=2$, we have horizons according to the values of $a$, with $n=1,a_{ext}=0.714410046190945m$ and $n=2,a_{ext}=0.4456300400812961661m$. 

The causal structure is given by:
(i) the cases $n=1$ and $a>a_{ext} \approx 0.714m$;
 $n=2$ and $a>a_{ext} \approx 0.446m$, are depicted in Fig.~\ref{figPenrose1};
(ii) $n=2$ and $a=a_{ext} \approx 0.446m$, in Fig.~\ref{figPenrose4};
(iii) $n=1$ and $a=a_{ext} \approx 0.714m$, in Fig.~\ref{figPenrose5};
(iv) $n=1$ and $a=a_{ext} \approx 0.714m$, in Fig.~\ref{figPenrose6}.

\begin{figure}[htb!]
\includegraphics[scale=0.6]{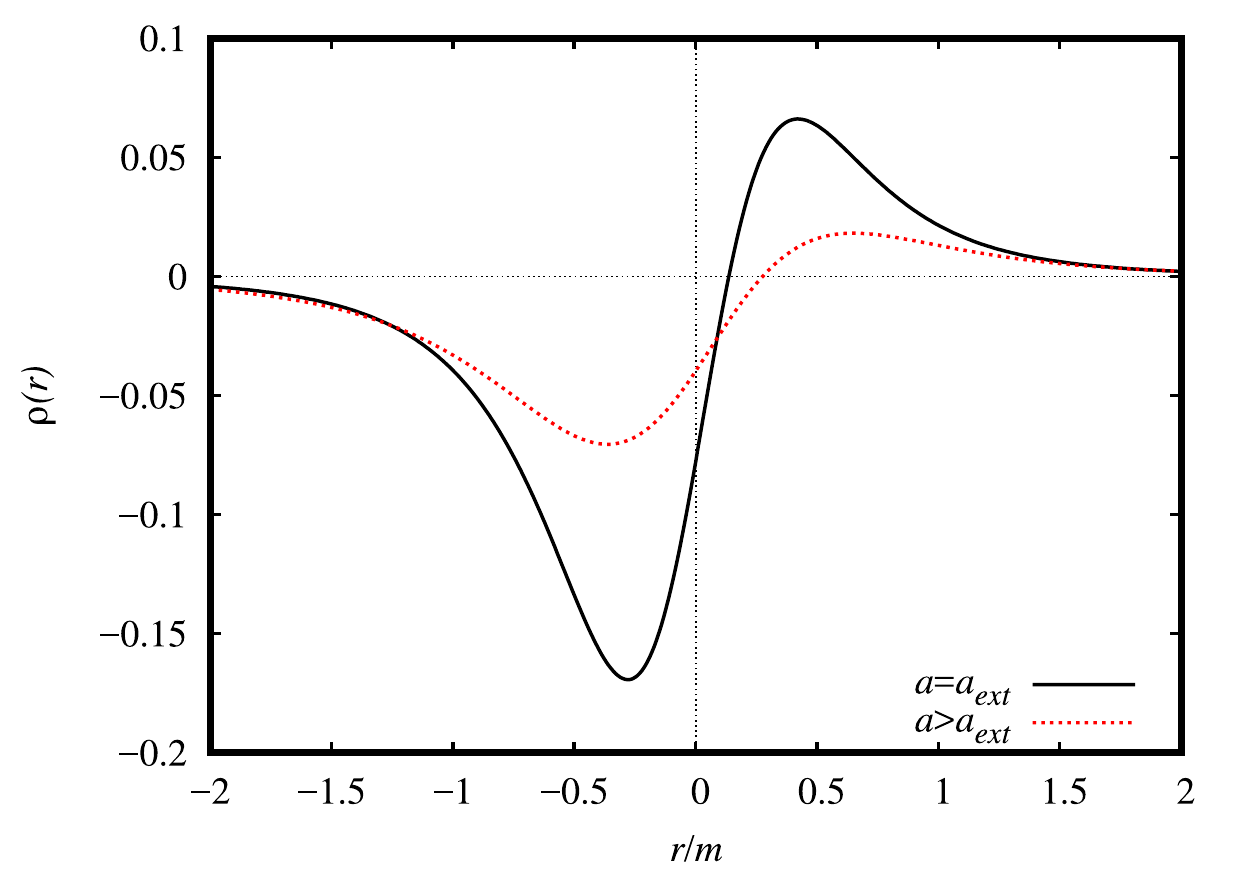}\qquad
	\includegraphics[scale=0.6]{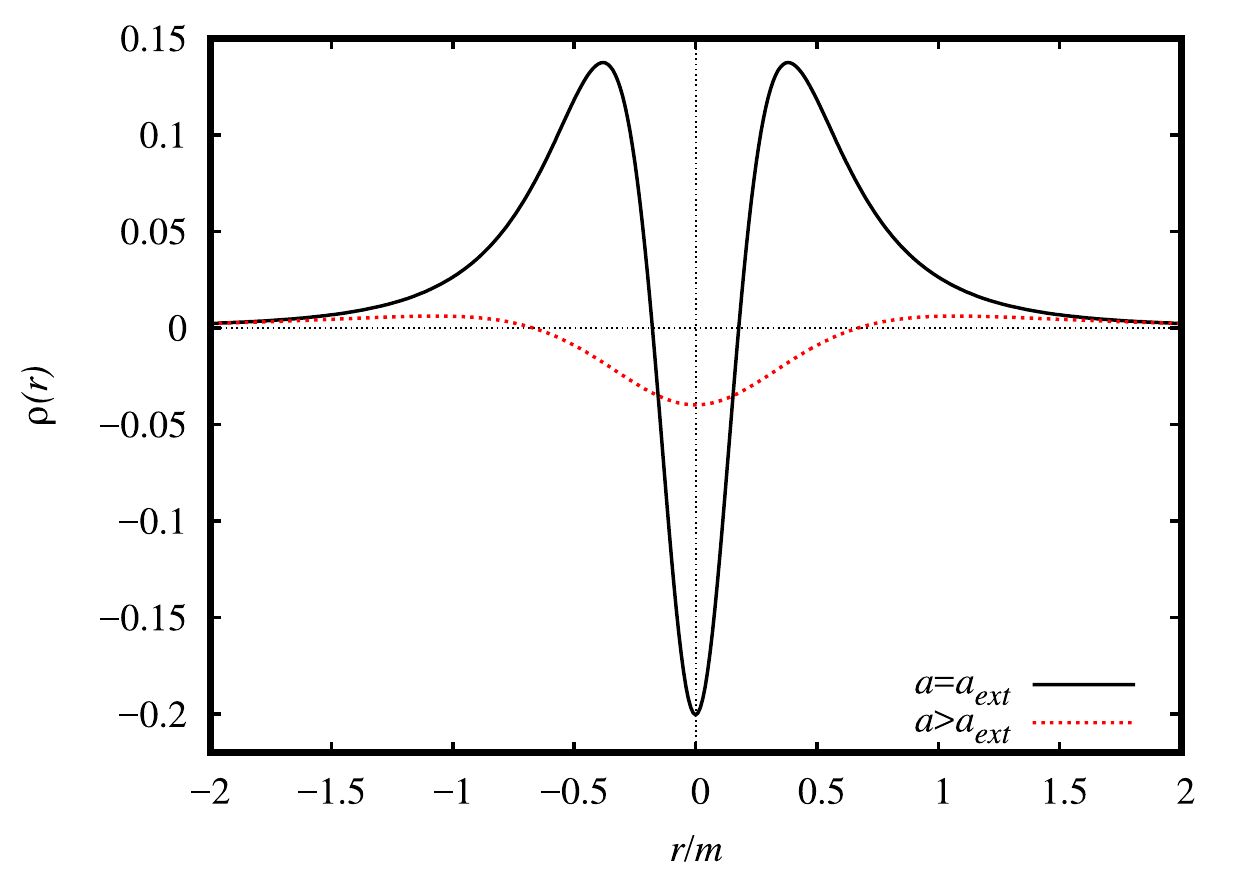}	\caption{Energy density for~\eqref{mod5} with $n=1,\kappa^2=8\pi$ (left) and $n=2, \kappa^2=8\pi$ (right).}\label{fig10}
\end{figure}

We notice that for odd $n$, the positive and negative regions of $r$ are not symmetric, contrary to the situation for $n$ even, where the regions are symmetric. In the case where $n=1$, the energy density is positive for $r>r_+$, and negative for $ -\infty <r <r_+ $, as we see in the left plot Fig.~\ref{fig10}. In the right plot of Fig.~\ref{fig10}, we have the case $n=2$, where the energy density is positive for $r>r_+$, and negative inside of the horizon.


The Hernandez--Misner--Sharp mass~\eqref{mass} for the model~\eqref{mod5} is given by
\begin{eqnarray}
M_\HMS(r)=\frac{a^2}{2 \sqrt{r^2+a^2}}+\frac{m \left(\frac{2}{\pi }\right)^n r^2 \arctan\left(\frac{r}{a}\right)^n}{r^2+a^2}\label{massmod5}\,.
\end{eqnarray}
For $n$ even the mass is always positive, and possesses the limits $\lim_{r\rightarrow \infty}M_\HMS(r)=m$ and $\lim_{r\rightarrow 0}M_\HMS(r)=a/2$.

\section{Conclusion}\label{S:conclusion}

The investigation of wormholes and regular compact objects in GR allows for the construction of multiple models that have the two main characteristics of the previous models, a throat and regularity (and possibly horizons); models which we call black-bounce. In this work we have presented two quite general theorems that guide two general characteristics of these models, the regularity of static space-times and the energy conditions for them. We note that curvature regularity can be readily verified 
by checking the condition that the Kretschmann scalar is always finite, and that we can easily characterize the situations under which the usual point-wise energy conditions are always violated for  spherically symmetric models with the metric~\eqref{ele}.
We have re-analysed the Simpson--Visser model, adding a new physical quantity to the discussion: the Hernandez--Misner--Sharp quasi-local mass, which is always positive. 

We present several new classes of black-bounce models that generalize the geometry of the original Simpson--Visser model. 
Two of them reduce to the original for a specific choice of parameters. All the models reduce to Schwarzschild for a suitable choice of parameters, are regular throughout the  spacetime, and have an area of the angular part of the metric that is always positive and non-zero. All models have positive Hernandez--Misner--Sharp mass. 
The first of these solutions, the model in subsection~\ref{SS:n=2+k=0}, has exactly the same characteristics as the Simpson--Visser model. 
The second, given in subsection~\ref{SS:n=1+k=2}, presents different characteristics, such as the possibility of having four horizons, two event horizons and two Cauchy horizons, and choosing the parameters properly, we can have an extreme case with only two event horizons. 
The causal structure of this solution with four horizons cannot be represented in a usual Penrose diagram, and the extreme case shows something new, namely, the bounce at $r=0$ is timelike, and can be traversed in both directions, both from $r>0$ to $r<0$ and from $r<0$ to $r>0$. 
This solution has the symmetry $a\leftrightarrow -a$ and $r\leftrightarrow -r$.
In subsection \ref{SS:zero} we analyse a model with zero energy density. Analysis of this model explicitly demonstrates that, while one can by construction force tolerable behaviour for the  $WEC_3$ constraint, one would still violate $NEC_1 = WEC_1 = SEC_1$ everywhere ``off-horizon''.
In the fourth solution, presented in subsection~\ref{SS:M(r)_1}, we also have the possibility of having multiple horizons, more than four, depending on the choice of parameters. 
The fifth solution, presented in subsection~\ref{SS:M(r)_2}, may present an asymmetry, such as two horizons on the positive part of the radial coordinate $r$ and none on the negative part. 
For the symmetric solution of this model, the energy condition $SEC_3=\rho-p_r-2p_t\geq 0$ may be satisfied for all $r$. This is a specific characteristic of black-bounce models, because in regular black holes with spherical symmetry this condition is always violated within the event horizon~\cite{Zaslavskii}. 
In the sixth model, presented in subsection~\ref{SS:M(r)_3}, we also have the possibility of asymmetry, but for the symmetrical model, the energy density is always positive outside the event horizon.

In conclusion, we have presented and analysed just some of the salient features of several models of new black-bounce geometries. 
We could also study geodesics, dynamic thin-shells, thermodynamics, the scattering and absorption of quantum fields, shadows (silhouettes) and quasinormal modes. 
These topics will be addressed appropriately in future work.

\appendix
\section{Regularity of  static spacetimes}\label{S:appendix}
In reference~\cite{Bronnikov:2012wsj}, Bronnikov and Rubin showed that for a spherically symmetric and static spacetime, finiteness of the Kretschmann scalar is enough to 
forbid a curvature singularity. We now state the following somewhat more general theorem that does not appeal to spherical symmetry. 
 
{\bf Theorem:} For any static spacetime, in the strictly static region, the Kretschmann scalar is positive semi-definite, being a sum of squares of the nonzero components $R^{\hat{a}\hat{b}}_{\ \ \hat{c}\hat{d}}$. Then if this scalar is finite, all the orthonormal components of the Riemann tensor must be finite.

{\it Proof:}
First, for any arbitrary spacetime in terms of any orthonormal basis, the Kretschmann scalar is
\begin{equation}
K=R_{\mu\nu\alpha\beta}R^{\mu\nu\alpha\beta}=R_{\hat{a}\hat{b}\hat{c}\hat{d}}R^{\hat{a}\hat{b}\hat{c}\hat{d}}.
\end{equation}
Now assuming only that one can distinguish space from time, split the indices into space and time,  $\hat{a}=(\hat{0},\hat{i})$, so that
\begin{equation}
K=R_{\hat{i}\hat{j}\hat{k}\hat{l}}R^{\hat{i}\hat{j}\hat{k}\hat{l}}
+4R_{\hat{0}\hat{i}\hat{j}\hat{k}}R^{\hat{0}\hat{i}\hat{j}\hat{k}}
+4R_{\hat{0}\hat{i}\hat{0}\hat{j}}R^{\hat{0}\hat{i}\hat{0}\hat{j}}+4R_{\hat{0}\hat{0}\hat{0}\hat{i}}R^{\hat{0}\hat{0}\hat{0}\hat{i}}+R_{\hat{0}\hat{0}\hat{0}\hat{0}}R^{\hat{0}\hat{0}\hat{0}\hat{0}}.
\end{equation}
But the last two terms vanish in view of the symmetries of the Riemann tensor, and so
\begin{equation}
K=R_{\hat{i}\hat{j}\hat{k}\hat{l}}R^{\hat{i}\hat{j}\hat{k}\hat{l}}
+4R_{\hat{0}\hat{i}\hat{j}\hat{k}}R^{\hat{0}\hat{i}\hat{j}\hat{k}}
+4R_{\hat{0}\hat{i}\hat{0}\hat{j}}R^{\hat{0}\hat{i}\hat{0}\hat{j}}.
\end{equation}
But since, in the strictly static region where the $t$ coordinate is timelike, we have  $g_{\hat{a}\hat{b}}=\eta_{\hat{a}\hat{b}}= {\rm diag} \{1,-1,-1,-1\}$, this reduces to
\begin{equation}
K=R_{\hat{i}\hat{j}\hat{k}\hat{l}}R_{\hat{i}\hat{j}\hat{k}\hat{l}}
-4R_{\hat{0}\hat{i}\hat{j}\hat{k}}R_{\hat{0}\hat{i}\hat{j}\hat{k}}
+4R_{\hat{0}\hat{i}\hat{0}\hat{j}}R_{\hat{0}\hat{i}\hat{0}\hat{j}}.
\end{equation}
Furthermore, in the strictly static region where the $t$ coordinate is timelike,
the 4-metric is block diagonalizable $g_{ab}=\left(N^2\right)\oplus (- g_{ij})$. More to the point the extrinsic curvature of the constant-$t$ spatial slices is then zero, and hence by the Gauss--Codazzi~\cite{Manfredo,wald,MTW} equations one has $R_{\hat{0}\hat{i}\hat{j}\hat{k}}=0$.

Thence as long as the spacetime is static we can split spacetime $\rightarrow$ space+time in such a manner that
\begin{equation}
K=R_{\hat{i}\hat{j}\hat{k}\hat{l}}R_{\hat{i}\hat{j}\hat{k}\hat{l}}
+4R_{\hat{0}\hat{i}\hat{0}\hat{j}}R_{\hat{0}\hat{i}\hat{0}\hat{j}}\geq 0.
\end{equation}
Consequently in any static spacetime if the Kretschmann scalar is finite then all the orthonormal components $R_{\hat{a}\hat{b}\hat{c}\hat{d}}$ of the Riemann tensor must be finite. Therefore, we can determine the regularity of a static space-time simply by checking if the Kretschmann scalar is finite.

Similar comments can be made about the Weyl tensor:
\begin{equation}
C_{\mu\nu\alpha\beta} C^{\mu\nu\alpha\beta}=
C_{\hat{i}\hat{j}\hat{k}\hat{l}}C_{\hat{i}\hat{j}\hat{k}\hat{l}}
-4C_{\hat{0}\hat{i}\hat{j}\hat{k}}C_{\hat{0}\hat{i}\hat{j}\hat{k}}
+4C_{\hat{0}\hat{i}\hat{0}\hat{j}}C_{\hat{0}\hat{i}\hat{0}\hat{j}}.
\end{equation}
But the static condition implies that {both} the 4-metric and the Ricci tensor are block diagonalizable. Thence both  $g_{ab}=\left(N^2\right)\oplus (- g_{ij})$ and $R_{ab} = R_{00}\oplus R_{ij}$. This now implies that in static spacetimes $C_{\hat{0}\hat{i}\hat{j}\hat{k}}=R_{\hat{0}\hat{i}\hat{j}\hat{k}}=0$. So as long as the spacetime is static we can split spacetime $\rightarrow$ space+time in such a manner that
\begin{equation}
C_{\mu\nu\alpha\beta} C^{\mu\nu\alpha\beta}=
C_{\hat{i}\hat{j}\hat{k}\hat{l}}C_{\hat{i}\hat{j}\hat{k}\hat{l}}
+4C_{\hat{0}\hat{i}\hat{0}\hat{j}}C_{\hat{0}\hat{i}\hat{0}\hat{j}}\geq 0.
\end{equation}
Consequently in any static spacetime if the Weyl scalar $C_{\mu\nu\alpha\beta} C^{\mu\nu\alpha\beta}$ is finite then all the orthonormal components $C_{\hat{a}\hat{b}\hat{c}\hat{d}}$ of the Weyl tensor must be finite.


\section*{Acknowledgements}
FSNL acknowledges support from the Funda\c{c}\~{a}o para a Ci\^{e}ncia e a Tecnologia (FCT) Scientific Employment Stimulus contract with reference CEECINST/00032/2018. FSNL also thanks funding from the FCT research grants No. UID/FIS/04434/2020, No. PTDC/FIS-OUT/29048/2017 and No. CERN/FIS-PAR/0037/2019.
MER  thanks Conselho Nacional de Desenvolvimento Cient\'ifico e Tecnol\'ogico - CNPq, Brazil  for partial financial support. This study was financed in part by the Coordena\c{c}\~{a}o de Aperfei\c{c}oamento de Pessoal de N\'ivel Superior - Brasil (CAPES) - Finance Code 001.
AS acknowledges financial support via a PhD Doctoral Scholarship provided by Victoria University of Wellington. AS is also indirectly supported by the Marsden fund, administered by the Royal Society of New Zealand.
MV was directly supported by the Marsden Fund, via a grant administered by the Royal Society of New Zealand.


\end{document}